\documentclass[%
superscriptaddress,
reprint,
amsmath,
amssymb,
aps,
longbibliography,
prb,
floatfix,
]{revtex4-2}

\usepackage{graphicx}% Include figure files
\usepackage{color}
\usepackage{dcolumn}% Align table columns on decimal point
\usepackage{bm}% bold math
\usepackage{hyperref}% add hypertext capabilities
\usepackage[normalem]{ulem} % To cross out text

\usepackage{physics}
\usepackage{float}

\DeclareMathOperator{\mueV}{\mu\text{eV}}

\begin{document}

\title{Protocols to measure the non-Abelian Berry phase\texorpdfstring{\\}{}by pumping a spin qubit through a quantum-dot loop}
\author{Baksa Kolok}
 \email{kolokba@edu.bme.hu}
\affiliation{%
 Department of Theoretical Physics, Institute of Physics, Budapest University of Technology and Economics, M\H{u}egyetem rkp. 3., H-1111 Budapest, Hungary
}%
\author{Andr{\'{a}}s P{\'{a}}lyi}%
\email{palyi.andras@ttk.bme.hu}
\affiliation{%
 Department of Theoretical Physics, Institute of Physics, Budapest University of Technology and Economics, M\H{u}egyetem rkp. 3., H-1111 Budapest, Hungary
}%
\affiliation{MTA-BME Quantum Dynamics and Correlations Research Group, M{\"{u}}egyetem rkp. 3., H-1111 Budapest, Hungary}

\date{\today}

\begin{abstract}
A quantum system constrained to a degenerate energy eigenspace can undergo a nontrival time evolution upon adiabatic driving, described by a non-Abelian Berry phase. 
This type of dynamics may provide logical gates in quantum computing that are robust against timing errors.
A strong candidate to realize such holonomic quantum gates is an electron or hole spin qubit trapped in a spin-orbit-coupled semiconductor, whose twofold Kramers degeneracy is protected by time-reversal symmetry.
Here, we propose and quantitatively analyze protocols to measure the non-Abelian Berry phase by pumping a spin qubit through a loop of quantum dots. 
One of these protocols allows to characterize the local internal Zeeman field directions in the dots of the loop. 
We expect a near-term realisation of these protocols, as all key elements have been already demonstrated in spin-qubit experiments.
These experiments would be important to assess the potential of holonomic quantum gates for spin-based quantum information processing.
\end{abstract}

\maketitle

\section{Introduction}

Recent demonstrations of coherent electron and hole spin control in semiconductor quantum dot arrays \cite{Hendrickx2021AProcessor, hendrickx2023sweetspot, Madzik2022PrecisionSilicon, Noiri2022FastSilicon, Xue2022QuantumThreshold, Philips2022UniversalSilicon} 
represent milestones toward a fault-tolerant quantum processor of spin qubits.
In such devices, spin-orbit interaction  often plays an important role.
For example, in the presence of a magnetic field, it enables resonant control of a spin qubit using an ac electric field \cite{Golovach2006Electric-dipole-inducedDots, Nowack2007CoherentFields, Crippa2018ElectricalQubits}.

\begin{figure}
    \centering
    \includegraphics[width=1\linewidth]{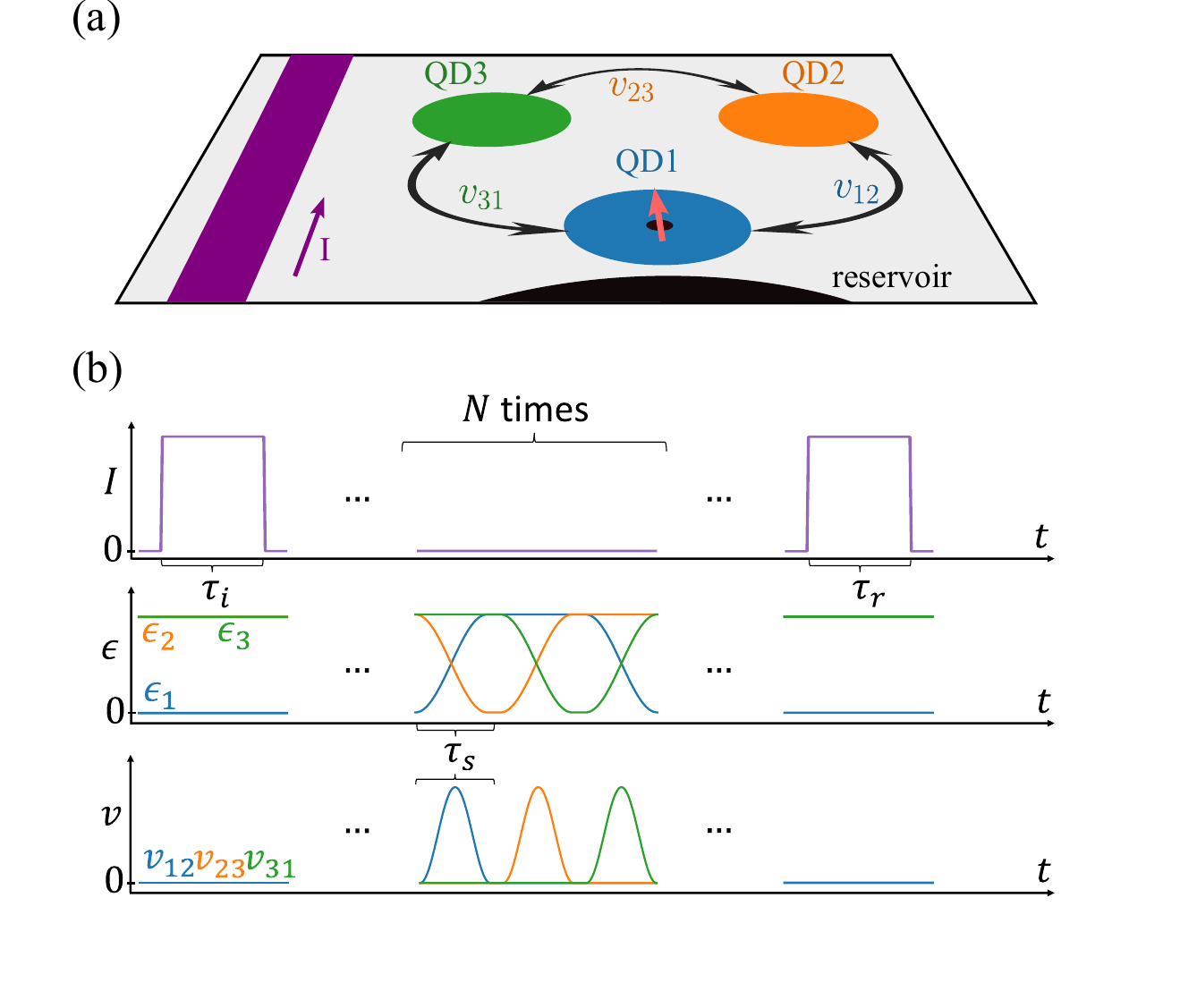}
    \caption{Protocol to measure the non-Abelian Berry phase by adiabatically pumping a spin qubit through a quantum-dot (QD) loop. 
    (a) Three QDs forming a loop. 
    A reservoir resides next to QD1, enabling Elzerman-type readout \cite{Elzerman2004Single-shotDot}. 
    A current ($I$) pulse through the wire (purple) induces magnetic field for initialization and readout.
    (b) Current ($I$) and gate-voltage ($\epsilon$, $v$) pulse sequence. 
    Gate voltages control QD on-site energies ($\epsilon$) and interdot tunneling energies ($v$). 
    See Sec.~\ref{sec:singletinit} for an alternative protocol that does not require pulsed magnetic fields, and relies on Pauli blockade for initialization and readout.
    }
    \label{fig:experiment}
\end{figure}

In the absence of a magnetic field, spin-orbit coupling does not break the twofold Kramers degeneracy of a spin qubit. 
However, spin-orbit does trigger qubit dynamics in the presence of time-dependent electric fields via geometric effects. 
On the one hand, if electric fields are intentionally used to move the qubit adiabatically along a closed loop in real space [such as in a loop of 3 quantum dots, as shown in Fig.~\ref{fig:experiment}(a)], then the qubit state undergoes a nontrivial time evolution in the two-dimensional Kramers-degenerate subspace, described by a path-dependent propagator that is often called the non-Abelian Berry phase \cite{Wilczek1984AppearanceSystems,Golovach2010HolonomicDots,San-Jose2008GeometricDecoherence, prem2023longitudinal}. 

Quantum gates based on such adiabatic control form the building blocks of holonomic quantum computing \cite{Zanardi1999HolonomicComputation}.
As simulations and experiments revealed, holonomic quantum computation may overcome the challenges of building a large-scale quantum computer due to its resistance to specific types of faults \cite{Sjoqvist2008AComputation}. Furthermore, as theoretical \cite{San-Jose2006GeometricalDots} and experimental \cite{hendrickx2023sweetspot} analyses showed, the effect of charge noise on semiconductor spin qubits becomes weaker as the Zeeman splitting of the qubit is decreased. According to these findings, holonomic quantum computing with such spin qubits, envisioned at zero external magnetic field, may have an advantage over standard resonant control that is carried out in a finite magnetic field.
On the other hand, the same mechanism implies that the interplay of spin-orbit coupling and electrical fluctuations lead to geometrical qubit dephasing \cite{San-Jose2008GeometricDecoherence, San-Jose2006GeometricalDots}.
To our knowledge, these geometric effects predicted for spin-orbit-coupled particles have not yet been demonstrated experimentally 
and the protocol proposed below is one step toward the realization of holonomic quantum computation in semiconductor quantum dot devices. 

The effect of the (Abelian) Berry phase \cite{Berry1984QuantalChanges} on non-degenerate levels has been observed in a variety of systems, 
from optical fibers \cite{Tomita1986ObservationFiber} and semiconductor rings \cite{Grbic2007Aharonov-BohmInteractions, Nagasawa2013ControlRings, Yau2002Aharonov-BohmPhase} to Bose condensates of ultracold atoms \cite{Lin2009SyntheticAtoms}.
In addition, geometrical dephasing due to this Berry phase have been measured in superconducting qubits \cite{Leek2007ObservationQubit, Berger2015MeasurementQubit}.

As for the non-Abelian Berry phase, which is the focus of our work, experimental evidence has been observed in nuclear magnetic resonance experiments \cite{Zwanziger1990Non-AbelianAxes}, and more recently in two-dimensional hole gases in semiconductors \cite{Li2016ManifestationSystem}, in Bose condensates \cite{Sugawa2021WilsonField, lv2023measurement}, in coupled photonic waveguide structures \cite{Kremer2019OptimalMetric} and in classical optical interference measurements using real space loops \cite{Yang2019SynthesisSpace}. 
However, to our knowledge, the non-Abelian Berry phase has not been experimentally studied in quantum dot systems. 
In contrast to the recent experiments on non-Abelian Berry phases, where degeneracy of the system is achieved by fine tuning of the system's parameters \cite{Sugawa2021WilsonField, lv2023measurement, Kremer2019OptimalMetric, Yang2019SynthesisSpace}, in a quantum dot system at zero magnetic field the energy degeneracy is due to time-reversal symmetry.
It is reasonable to expect that this symmetry-based protection of the degeneracy is more robust than protection based on fine tuning. 
Hence, we consider spin qubits in the presence of spin-orbit coupling and absence of magnetic field as an outstanding experimental platform to measure non-Abelian Berry phase effects.

Here, we propose and theoretically analyse a protocol to measure the non-Abelian Berry phase by adiabatically pumping a spin qubit through a quantum-dot loop. 
Our main protocol, see Fig.~\ref{fig:experiment}, incorporates dc-pulsed magnetic fields for initialisation and readout, and qubit shuttling at zero magnetic field. 
We also propose a second version of the protocol to infer the key spin-orbit parameters of the multi-dot system, and a third version which uses Pauli-blockade initialisation and readout, and requires neither the magnetic field dc-pulse nor tunnelling to a reservoir. 
Furthermore, we describe variants for explicit demonstration of the non-Abelian (i.e., non-commutative) nature of the adiabatic geometric quantum gates. 
We anticipate that hole spin qubits in two-dimensional (2D) arrays of germanium quantum dots \cite{Hendrickx2021AProcessor, borsoi2022shared, hendrickx2023sweetspot, vanRiggelen2022PhaseQubits, wang2022probing} are prominent candidates to carry out these experiments, due to their high device quality, planar 2D layout, strong spin-orbit interaction, and weak hyperfine effects.
Note that single-electron shuttling experiments have been carried out in GaAs-based devices \cite{Flentje2017CoherentSpins, Mortemousque2021EnhancedDots}; however, there hyperfine interaction is comparatively strong, hindering spin qubit functionalities, and potentially masking the non-Abelian Berry phase effect.

The rest of the paper is organized as follows. 
In Sec.~\ref{sec:looppumping}, 
we present a minimal model describing the non-Abelian Berry phase propagator of a spin qubit as it is pumped through a quantum-dot loop.
In Sec.~\ref{sec:NABP_meas}, we propose a measurement protocol to observe the non-Abelian Berry phase.
In Sec.~\ref{sec:parametrisation}, we show that the interdot pseudospin-non-conserving tunneling parameters and the angles characterizing the local Zeeman fields can be inferred via a similar measurement protocol. 
We discuss an alternative protocol based on Pauli-blockade initialization and readout, as well as risks, challenges and further opportunities related to the proposed experiments, in Sec.~\ref{sec:discussion}. 
We draw our conclusions in Sec.~\ref{sec:conclusion}.

\section{Adiabatic charge pumping through a quantum dot loop in a minimal model} \label{sec:looppumping}

In this section, we introduce a minimal model to describe the non-Abelian Berry phase in a loop of three quantum dots.

\subsection{Setup, model, and the non-Abelian Berry phase} \label{sec:looppumping_results}

Consider a particle (an electron or a hole) in a quantum dot loop, subject to spin-orbit interaction.
The setup is shown in Fig.~\ref{fig:experiment}(a).
Assume that the qubit is initialized in a specific state in the two-dimensional Kramers-degenerate subspace associated to the lowest-lying orbital of QD1. 
By slowly changing the plunger and barrier gate voltages, as depicted in Fig.~\ref{fig:experiment}(b), the qubit is adiabatically pumped through the loop.
As a result of being moved in the presence of spin-orbit interaction, the qubit state is rotated by the end of a full cycle.
This qubit rotation can be phrased in terms of a non-Abelian Berry phase, which we describe as follows.
 
We take into account the ground-state orbitals of each QD, and the corresponding Kramers doublets which we call pseudospin doublets. 
Hence, with three dots, the Hilbert space is six dimensional. 
As long as we work with arbitrary pseudospin basis states, the model Hamiltonian of this setup reads
\begin{subequations}
\begin{eqnarray}
    \label{eq:dotloop_tilde}
    H &=& H_{\textrm{on-site}} + H_{\textrm{tun}} + H_{\textrm{stun}}, \\
    H_{\textrm{on-site}} &=& \epsilon_1 \tau_1 + \epsilon_2 \tau_2 + \epsilon_3 \tau_3, \\
    H_{\textrm{tun}} &=& \tilde{v}_{12} \tau^{x}_{12} + \tilde{v}_{23} \tau^{x}_{23} + \tilde{v}_{31} \tau^{x}_{31}, \label{eq:tildetun} \\
    H_{\textrm{stun}} &=& \hbar  \boldsymbol{\tilde{\Omega}}_{12} \cdot \boldsymbol{\tilde{\sigma}} \otimes \tau^{y}_{12} + \hbar  \boldsymbol{\tilde{\Omega}}_{23} \cdot \boldsymbol{\tilde{\sigma}} \otimes \tau^{y}_{23} +\nonumber \\ & & \quad +\, \hbar \boldsymbol{\tilde{\Omega}}_{31} \cdot \boldsymbol{\tilde{\sigma}} \otimes \tau^{y}_{31}, \label{eq:tildestun}
\end{eqnarray}
\label{eq:hamiltonian}
\end{subequations}
where $\tilde{\sigma}_{\alpha}$ is the $\alpha$-th Pauli operator acting on the pseudospin degree of freedom for $\alpha \in \{ x, y, z\}$, $\boldsymbol{\tilde{\sigma}}$ is the vector of $\tilde{\sigma}_\alpha$, $\tau_k$ with $k\in \{1,2, 3\}$ is the projection on the $k$-th dot, and $\tau_{jk}^\alpha$ with $j,k\in \{1,2, 3\}$ is the Pauli $\alpha$ operator between the dots $j$ and $k$. 

The matrix forms of these operators can be expressed using the basis vectors representing wave functions localized on quantum dot $k$.
These basis vectors are denoted as $\ket*{k, \tilde{\Downarrow}}$ and $\ket*{k, \Tilde{\Uparrow}}$, and they are assumed to be Kramers pairs, i.e., 
$\ket*{k, \tilde{\Downarrow}} = \mathcal{T} \ket*{k, \Tilde{\Uparrow}}$ with $\mathcal{T}$ being the time-reversal operator.
The operators in Eq.~\eqref{eq:hamiltonian} read, for example, as $\tau_k = \ket*{k, \tilde{\Downarrow}}\bra*{k, \tilde{\Downarrow}} + \ket*{k, \tilde{\Uparrow}}\bra*{k,\tilde{\Uparrow}}$, $\tau^x_{jk} = \ket*{j}\bra*{k} + \ket*{k}\bra*{j}$ and $\tau^y_{jk} = -i\ket*{j}\bra*{k} + i\ket*{k}\bra*{j}$. 
Tunneling between the $j$-th and the $k$-th dot is described by four parameters (for each QD pair: one scalar and three coordinates of a vector):  $\tilde{v}_{jk}$ is the pseudospin-conserving tunneling, and $\tilde{\boldsymbol{\Omega}}_{jk}$ is the vector describing the pseudospin-non-conserving tunneling. 
The pseudospin-non-conserving tunneling terms in the Hamiltonian  are induced by spin-orbit coupling \cite{Danon2009PauliCoupling}.

\begin{figure}
    \centering
    \includegraphics[width=0.8\linewidth]{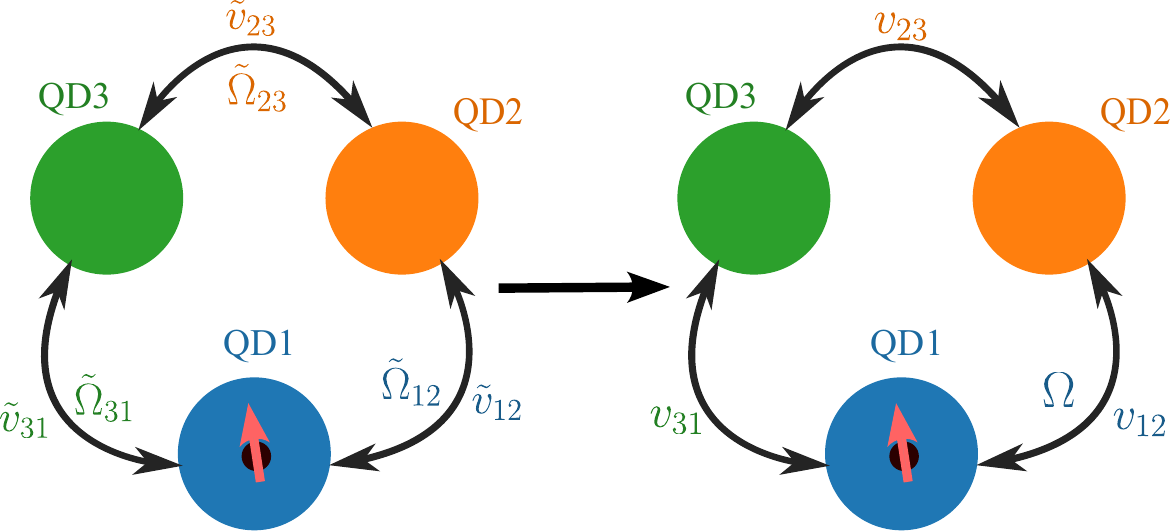}
    \caption{A convenient gauge transformation for calculating the non-Abelian Berry phase. 
    The pseudospin-non-conserving tunneling matrix elements between the dot pairs (QD2,QD3) and (QD3,QD1) are eliminated by the gauge transformation, which in turn transforms the pseudospin-non-conserving tunneling matrix elements between the dot pair (QD1,QD2) into  $\boldsymbol{\Omega}$.
    By a further simultaneous pseudospin rotation on all three dots, $\boldsymbol{\Omega} = \Omega(0,0,1)$ can be achieved.}
    \label{fig:dotloop_gauge}
\end{figure}

Since the Hamiltonian $H$ describes a single fermion in the presence of time-reversal symmetry (no magnetic field), the energy levels of $H$ are twofold degenerate, known as Kramers degeneracy. 
Furthermore, the matrix form of the Hamiltonian depends on the choice of the localized basis vectors, which can be considered as a gauge degree of freedom. 
With an appropriate gauge transformation, we can simplify the Hamiltonian, as illustrated in Fig.~\ref{fig:dotloop_gauge}: 
(i) Two of the three tunneling Hamiltonian terms, e.g., those between QD pairs (1,3) and (2,3), can be transformed to a pseudospin-conserving form.
(ii) The pseudospin-non-conserving tunneling parameters of the third QD pair (1,2), obtained after this gauge transformation, form a vector with a single positive component,  $\boldsymbol{\Omega} = (0,0,\Omega)$.

In fact, after this gauge transformation, the tunneling terms 
$H_{\textrm{tun}}$ and $H_\textrm{stun}$
transform into the following form:
\begin{subequations}
\label{eq:tunaftergauge}
\begin{eqnarray}
    \label{eq:tun_preserving}
    H_\textrm{tun} &=& v_{12} \tau^{x}_{12} + v_{23} \tau^{x}_{23} + v_{31} \tau^{x}_{31}, \\
    \label{eq:stun_preserving}
    H_\textrm{stun} &=& \hbar \Omega \sigma_z \otimes \tau^{y}_{12}. \label{eq:stunaftergauge}
\end{eqnarray}
\end{subequations}
The parameters of these terms can be expressed with the old parameters, see App.~\ref{app:gauge} for the formulas and the detailed derivation. 
The Pauli matrices in Eq.~\eqref{eq:tunaftergauge} are expressed in the new basis, denoted as $\ket{k, \Downarrow}, \ket{k, \Uparrow}$. 
Note that the above gauge specification still leaves a gauge freedom, in the sense that a further global pseudospin rotation around the z axis leaves the Hamiltonian unchanged.

In our chosen gauge, 
the unitary qubit propagator $U$ (a.k.a. `the non-Abelian Berry phase' or a `holonomic quantum gate') corresponding to a full adiabatic pumping cycle $\text{QD1}\to \text{QD2} \to \text{QD3} \to \text{QD1}$ 
of the qubit through the loop reads
\begin{equation} \label{eq:nonAB}
    U = \frac{1}{\sqrt{v_{12}^2 + \hbar^2\Omega^2}}\begin{pmatrix}
    v_{12} + i\hbar\Omega & 0 \\
    0 & v_{12} - i\hbar\Omega
    \end{pmatrix} = e^{i\frac{\theta_\text{so}}{2} \sigma_z},
\end{equation}
where $\theta_\text{so} = 2\arctan \left(\frac{\hbar \Omega}{v_{12}}\right)$, and the basis ordering is $\ket{1,\Uparrow}$, $\ket{1,\Downarrow}$.
This unitary $U$ is a pseudospin rotation around the $z$ axis with angle $\theta_\text{so}$.
We provide the derivation of this result in the next subsection.

\subsection{Derivation of Eq.~(\ref{eq:nonAB})}

The  propagator $U_\text{cycle}$ describing the time evolution of the qubit through the complete adiabatic cycle $\text{QD1}\to \text{QD2} \to \text{QD3} \to \text{QD1}$ is a $6 \times 6$ matrix, which can be represented with the basis ordering $\ket{1,\Uparrow},\ket{1,\Downarrow},\ket{2,\Uparrow}, \dots, \ket{3,\Downarrow}$.
This cycle propagator can be expressed using the $6\times 6$ propagators 
$W_{21}$, $W_{32}$ and $W_{13}$
of each tunneling step as
\begin{equation} \label{eq:shuttling_steps}
    U_\text{cycle} = 
    W_{13} W_{32} W_{21}.
\end{equation}

We calculate the propagator $W_{21}$, which corresponds to the first shuttling step, using the block-diagonal unitary transformation $V = \text{diag}(I, U^\dagger, I)$, where $U$ is the unitary defined in Eq.~\eqref{eq:nonAB}. 
This transformation of the Hamiltonian renders the tunneling between QD1 and QD2 pseudospin-conserving. 
Although this transformation renders the tunneling between QD2 and QD3 pseudospin-non-conserving, this does not complicate the calculation as tunneling between QD2 and QD3 is suppressed in this shuttling step (see Fig.~\ref{fig:experiment}b, showing that the supports of $v_{12}(t)$ and $v_{23}(t)$ are disjoint). 
Thus, the transformed Hamiltonian is trivial for the pseudospin degree of freedom.

Together with the adiabatic nature of the tunneling dynamics, these imply that the propagator for the first tunneling step has the following block-matrix form:
\begin{equation}
    \tilde{W}_{21} = \begin{pmatrix}
    0 & e^{-i\phi_{21}^+} & 0\\
    e^{-i\phi_{21}^-} & 0 & 0\\
    0 & 0 & e^{-i\epsilon_3\tau_\textrm{s}}
    \end{pmatrix}.
\end{equation}
Here, $\phi_{21}^\pm$ are dynamical phases  depending on the actual pulse shapes and timings, and $\tau_\textrm{s}$ is the time of the shuttling. 
Transforming this propagator back, we obtain:
\begin{equation}
    W_{21} = V^{\dagger}\Tilde{W}_{21}V = \begin{pmatrix}
    0 & e^{-i\phi_{21}^+}U^\dagger & 0\\
    e^{-i\phi_{21}^-}U & 0 & 0\\
    0 & 0 & e^{-i\epsilon_3\tau_\textrm{s}}
    \end{pmatrix}.
\end{equation}

For the other two shuttling steps, the tunneling is pseudospin-conserving, therefore the propagators have the following form:
\begin{subequations}
\begin{eqnarray}
    W_{32} = \begin{pmatrix}
    e^{-i\epsilon_1\tau_\textrm{s}} & 0 & 0\\ 
    0 & 0 & e^{-i\phi_{32}^+}\\
    0 & e^{-i\phi_{32}^-} & 0
    \end{pmatrix},\\
    W_{13} = \begin{pmatrix} 
    0 & 0 & e^{-i\phi_{13}^-}\\
    0 & e^{-i\epsilon_2\tau_\textrm{s}} & 0\\
    e^{-i\phi_{13}^+} & 0 & 0
    \end{pmatrix},
\end{eqnarray}
\end{subequations}
where $\phi_{32}^\pm$ and $\phi_{13}^\pm$ are dynamical phases. 

Multiplying together the three propagators as in Eq.~\eqref{eq:shuttling_steps}, we arrive to the cycle propagator that reads
\begin{equation} \label{eq:Ucycle}
U_\text{cycle} = 
\begin{pmatrix}
        e^{-i\phi_{1}} U & 0 & 0 \\ 
        0 & 0 & e^{-i\phi_2} \\
        0 & e^{-i\phi_{3}} U^{\dagger} & 0
    \end{pmatrix}.
\end{equation}
Here, the phases are dynamical and could be expressed from the dynamical phases described above. 
As long as the qubit is localized on the first dot at the start of the process, the relevant part of $U_\text{cycle}$ is its top left $2\times 2$ block with $U$ being defined in Eq.~\eqref{eq:nonAB}, and the dynamical phases have no physical significance. 

\section{Measurement protocol for the non-Abelian Berry phase} \label{sec:NABP_meas}

In the previous section, we derived a formula, Eq.~\eqref{eq:nonAB}, for the non-Abelian Berry phase that characterizes an adiabatic pumping cycle of a single particle in a loop of three quantum dots. 
In general, the non-Abelian Berry phase of such a spin-$\frac{1}{2}$ particle is a 2$\times$2 unitary transformation. 
This unitary transformation corresponds to a rotation on the Bloch-sphere, therefore it can be described with an angle and an axis.

In this section, we propose a measurement to determine the angle and the axis corresponding to the non-Abelian Berry phase induced by an adiabatic pumping cycle in the quantum dot loop. 
In fact, the angle is a gauge-invariant physical quantity.
The axis, however, is a gauge-dependent quantity; below we describe how the axis is characterized by our measurement protocol.

We consider a single particle in the three-dot loop as described above, and shown in Fig.~\ref{fig:experiment}a.
The setup includes a metallic wire (purple stripe in Fig.~\ref{fig:experiment}a) that can host current dc pulses (purple solid line in Fig.~\ref{fig:experiment}b), and a particle reservoir (black in Fig.~\ref{fig:experiment}a) that is used for Elzerman-type qubit readout\cite{Elzerman2004Single-shotDot}. 

The protocol starts with having a single particle in QD1, and having no particles in QD2 and QD3.
As the first step, a current dc pulse is applied on the wire, to produce a dc magnetic field pulse that is felt by the particle in QD1 \cite{Koppens2006DrivenDot, Pla2013High-fidelitySilicon}.
The duration of the dc pulse is long enough such that the pseudospin of the particle in QD1 can thermalize, i.e., the pseudospin is initialised with a sufficiently high fidelity.
After switching off the current (and hence the magnetic field), the particle is adiabatically pumped through the loop via the sequence $\text{QD1} \to \text{QD2} \to \text{QD3} \to \text{QD1}$, to induce the non-Abelian Berry phase. 
This step might be repeated $N$ times, which amounts to performing the same geometric gate $N$ times. 
Finally, the qubit is read out in QD1 by Elzerman readout, utilizing a second current dc pulse.

Let us denote the probability of reading out the qubit in its excited state after the $N$th cycle as $P^{(\text{m})}_{1,\textrm{e},N}$. 
Below, we argue that the angle and the axis of the non-Abelian Berry phase can be characterized from  the experimental data of this readout probability $P^{(\text{m})}_{1,\textrm{e},N}$.

For this analysis, we first describe the time duration over which the initial current dc pulse is switched on, creating a magnetic field $\mathbf{B}$. 
For simplicity, we assume that this magnetic field is homogeneous over the volume of QD1. 
The Zeeman term induced by the magnetic field, expressed in the same general gauge as used in Eq.~\eqref{eq:dotloop_tilde}, takes the following form:
\begin{equation} \label{eq:zeeman}
    H_\textrm{Zeeman} = \sum_{k=1}^3 \frac{1}{2} \mu_\textrm{B} \boldsymbol{\Tilde{\sigma}} \cdot \mathbf{\tilde{g}}_k \mathbf{B} \otimes \tau_k,
\end{equation}
where $\mathbf{\tilde{g}}_k$ is the $g$-tensor of the $k$-th dot (a real, not necessarily symmetric 3$\times$3 matrix), and $\mu_\textrm{B}$ is the Bohr magneton.
We also introduce the notion of the local Zeeman vectors: $\hbar \boldsymbol{\tilde{\omega}}_k = \mu_\textrm{B}\mathbf{\tilde{g}}_k\mathbf{B}$. 

Let us transform the Hamiltonian $H$, which is now supplemented by the Zeeman term $H_\text{Zeeman}$, to the same gauge as it is in Eqs.~\eqref{eq:tun_preserving} and \eqref{eq:stun_preserving}. 
This means a rotation of the $g$-tensors (see App.~\ref{app:gaugeZeeman} for details). 
As it was mentioned before, the gauge specification in Sec.~\ref{sec:looppumping_results} still leaves a residual gauge freedom, i.e., we can further specify the gauge with a global pseudospin rotation that affects the Zeeman term only but leaves the other three terms in $H$ invariant. 
With the appropriate choice of the rotation, we rotate the local Zeeman vector of the first dot into the $xz$ plane with $x>0$.
Effectively, this rotation transforms the $g$-tensors of the Zeeman term. 

We denote the $g$-tensors after the transformations as $\mathbf{g}_k$. 
Then, the Hamiltonian of the system in that gauge is the following:
\begin{subequations} \label{eqs:conserving_gauge}
\begin{align}
    \label{eq:Hdotloop}
    H &= H_\textrm{on-site}  + H_\textrm{tun} + H_\textrm{stun} + H_\textrm{Zeeman}, \\
    \label{eq:dotloop_onsite}
    H_\textrm{on-site} &= \epsilon_1 \tau_1 + \epsilon_2 \tau_2 + \epsilon_3 \tau_3, \\
    \label{eq:dotloop_tun}
    H_\textrm{tun} &= v_{12} \tau^{x}_{12} + v_{23} \tau^{x}_{23} + v_{31} \tau^{x}_{31}, \\
    \label{eq:dotloop_stun}
    H_\textrm{stun} &= \hbar \Omega \sigma_z \otimes \tau^{y}_{12} \\
    \label{eq:dotloop_Zeeman}
    H_\textrm{Zeeman} &= \frac{1}{2}\mu_\textrm{B} B \sum_{k=1}^{3} g_k(\mathbf{n}_k \cdot \boldsymbol{\sigma}) \otimes \tau_k.
\end{align}    
\end{subequations}
where $g_k = \frac{\left|\mathbf{g}_k \mathbf{B}\right|}{B}$ is the $g$-factor of the $k$-th dot corresponding to the given direction of the magnetic field and $\mathbf{n}_k = (\sin\theta_k\cos\phi_k, \sin\theta_k\sin\phi_k, \cos\theta_k) = \frac{\mathbf{g}_k \mathbf{B}}{g_kB}$ is the unit vector in the direction of the local Zeeman field on the $k$-th dot. Moreover, $\phi_1 = 0$, because, as we mentioned above, the local Zeeman vector in QD1 is in the $xz$ plane with $x>0$.

The qubit ground state and excited state in QD1 are expressed in the basis $\ket*{1,\Uparrow}, \ket*{1, \Downarrow}$ as
\begin{align}
    \ket{1, \text{g}} = \begin{pmatrix}
    \sin\frac{\theta_1}{2} \\[2pt]
    -\cos\frac{\theta_1}{2}
    \end{pmatrix}, \ 
    \ket{1, \text{e}} = \begin{pmatrix}
    \cos\frac{\theta_1}{2} \\[2pt]
    \sin\frac{\theta_1}{2}
    \end{pmatrix}.
\end{align}
Therefore, the thermal equilibrium state of the system after the initialisation is
\begin{equation} \label{eq:thermastate1}
    \rho_1 = \frac{1}{1 + e^{-\frac{\hbar\omega_1}{k_BT}}}  \left(\ket{1,\text{g}}\bra{1,\text{g}} + e^{-\frac{\hbar\omega_{1}}{k_BT}}\ket{1,\text{e}}\bra{1,\text{e}}\right),
\end{equation}
with $\omega_k = |\boldsymbol{\tilde{\omega}}_k|$.
After $N$ pumping cycles, the pseudospin excited state occupation in QD1 yields:
\begin{align} \label{eq:occupationNloop}
    P_{1,\textrm{e},N} &= \bra{1, \text{e}} U_\text{cycle}^{N} \rho_1 U_\text{cycle}^{\dagger N} \ket{1,\text{e}} = \nonumber\\ 
    &= a (\theta_1, T, \omega_1) - b (\theta_1, T, \omega_1) \cos(N\theta_\text{so})
\end{align}
where $\theta_\text{so} = 2\arctan\left(\frac{\hbar\Omega}{v_{12}}\right)$, and we introduced the following quantities:
\begin{subequations} \label{eqs:rawparameters}
\begin{align}
    a (\theta_1, T, \omega_1) &= \frac{1}{2} - \frac{\tanh{\left(\frac{\hbar\omega_1}{k_BT}\right)} \cos^2\theta_1}{2},\\
    b (\theta_1, T, \omega_1) &= \frac{\tanh{\left(\frac{\hbar\omega_1}{k_BT}\right)}\sin^2\theta_1}{2}. \label{eq:ampe}
\end{align}
\end{subequations}

\begin{figure}
    \centering
    \includegraphics[width=\linewidth]{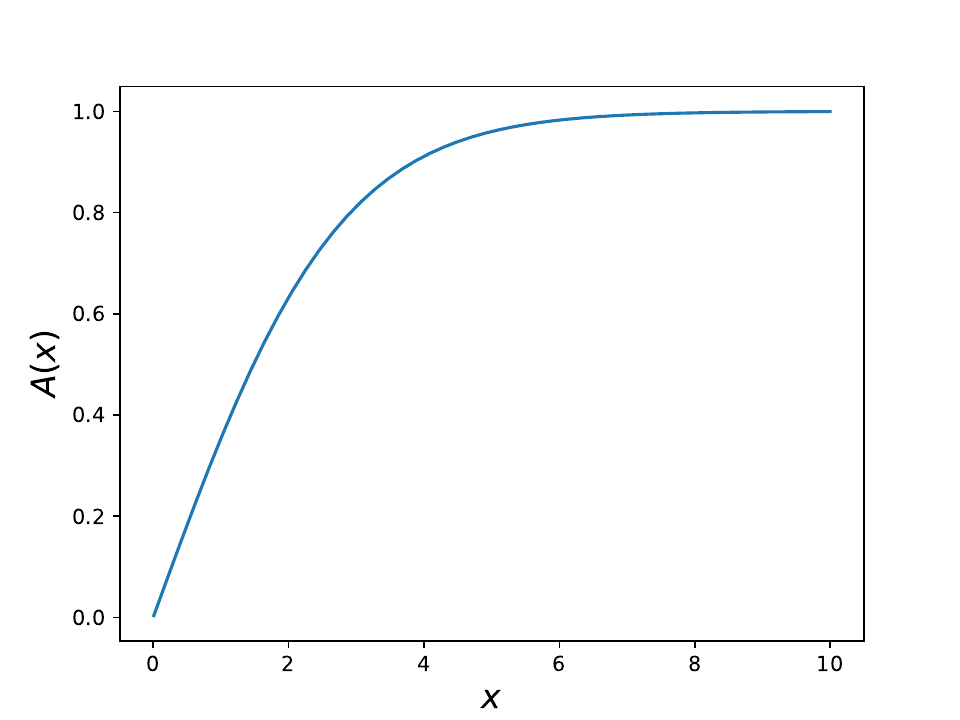}
    \caption{The function $A(x)$ defined in Eq.~\eqref{eq:damping}, which reduces the signal contrast due to the readout error caused by the thermal broadening of Fermi-Dirac distribution of particles in the reservoir.}
    \label{fig:readout}
\end{figure}

However, due to the finite temperature of the reservoir, the Elzerman-type readout does not yield the occupation probabilities. The actual measurement outcome ($P^{(\text{m})}_{1,\textrm{e},N}$) has the same form as \eqref{eq:occupationNloop} but with modified parameters:
\begin{align} \label{eq:measurementoutcomeNloop}
    P^{(\text{m})}_{1,\textrm{e},N} &= a^{(\text{m})} (\theta_1, T, \omega_1) - b^{(\text{m})} (\theta_1, T, \omega_1) \cos(N\theta_\text{so}).
\end{align}
This means that the contrast of the oscillating signal is reduced as
\begin{subequations} \label{eqs:contrastmod}
\begin{align}
    b&^{(\text{m})}(\theta_1, T, \omega_1) = A\left(\frac{\hbar\omega_1}{2k_\textrm{B}T}\right)\frac{\tanh{\left(\frac{\hbar\omega_1}{k_BT}\right)}\sin^2\theta_1}{2}, \label{eq:amp}\\
    A&(x) = \exp\left(\frac{x}{1 - e^{x}}\right) - \exp\left(\frac{x}{e^{-x} - 1}\right).
    \label{eq:damping}
\end{align}
\end{subequations}
We illustrated the function $A(x)$ in Fig.~\ref{fig:readout} to show how the signal contrast is modified if the Zeeman splitting is comparable to the temperature of the reservoir.
The offset parameter $a^{(\text{m})} (\theta_1, T, \omega_1)$  is also changed due to the imperfect readout, see the details of the calculation in App.~\ref{app:readout}. 

\begin{figure}[t]
    \begin{flushleft}
    \includegraphics[width=.95\linewidth]{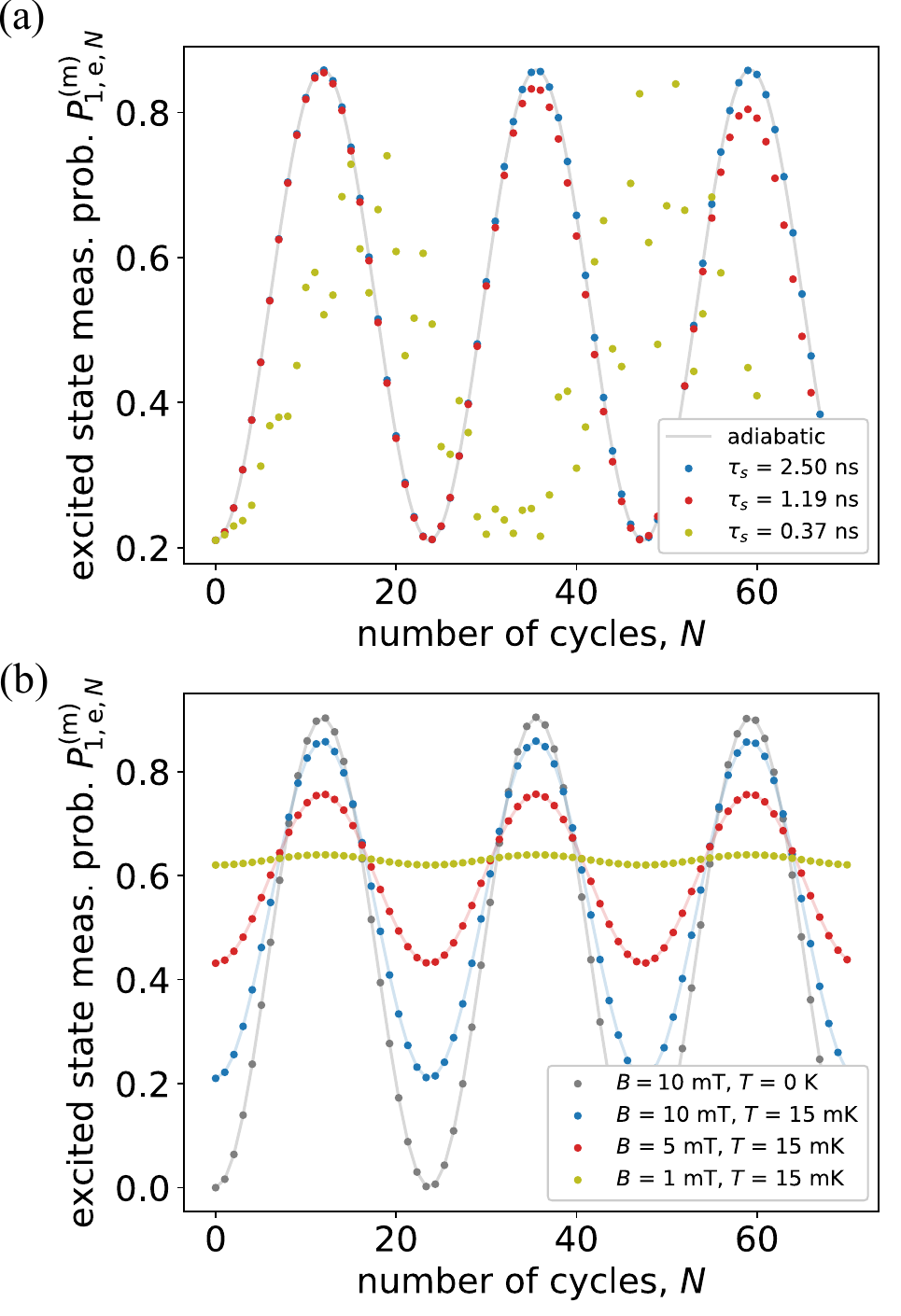}
    \end{flushleft}
    \caption{Effect of the non-Abelian Berry phase on pumping a spin-qubit through a quantum-dot loop. 
    (a) Simulation results of the measured excited state probability after $N$ pumping cycles.
    Gray solid line shows the result obtained from ideal adiabatic evolution, Eq.~\eqref{eq:measurementoutcomeNloop}. 
    Further data sets correspond to different shuttling times, see inset. 
    For slow shuttling, $\tau_\textrm{s} = 2.5$ ns, the simulation result matches the solid analytical result. 
    For faster braiding, $\tau_\textrm{s} \lesssim 2$ ns, the evolution is not adiabatic anymore. 
    Parameters: $\epsilon_k = 30 \mueV$ $v_{12} = 30 \mueV,\,v_{23} = 28 \mueV,\,v_{31} = 32 \mueV,\,\hbar\Omega = 4\mueV, \, T = 15\,\text{mK},\, B = 10 \, \text{mT},\,g_1 = 11,\,\theta_1 = 72$\textdegree, where the parameters $\epsilon_k$ and $v_{jk}$ are the amplitudes of the applied signals for the shuttling pulses shown in Fig.~\ref{fig:experiment}b. 
    Pulse shapes are sine for $\epsilon_k$-s and sine-squared for $v_{jk}$-s.
    (b) Analytical result Eq.~\eqref{eq:measurementoutcomeNloop} for the measured excited state probability for different magnetic field strengths used for the initialisation and different temperatures. 
    Parameters as above. 
    Gray solid line is the ideal case when $T = 0$ K, i.e., when initialisation and  readout are perfect.}
    \label{fig:NABP_diabatic}
\end{figure}

Equation \eqref{eq:measurementoutcomeNloop} is the central result of this work.
It is an approximate result, valid in the adiabatic limit.
To illustrate its relation to experiments, where the adiabatic approximation might break down, in Fig.~\ref{fig:NABP_diabatic}a we show simulation results obtained by solving the time-dependent Schr\"odinger equation numerically.

In the simulation, one shuttling step of the particle from the $j$th QD to the $k$th QD is induced by the following pulses:
\begin{subequations}
\begin{eqnarray}
    \epsilon_j (t) = \epsilon_0 \sin^2 \frac{t\pi}{2\tau_s}, \\ 
    \epsilon_k (t) = \epsilon_0 \cos^2 \frac{t\pi}{2\tau_s}, \\
    v_{jk} = v_0\sin^2 \frac{t\pi}{\tau_s}.
\end{eqnarray}
\end{subequations}
Furthermore, for the shuttling between QD1 and QD2 (where the tunneling is pseudospin-non-conserving), the parameter $\Omega$ is varied in time in the same way as $v_{jk}$. We used the above time-dependent parameters to define the time-dependent Hamiltonian, and numerically solved the Schr\"odinger equation to obtain the time evolution of the thermal state defined in Eq.~\eqref{eq:thermastate1} as the initial state.

We emphasize that whenever the adiabatic condition holds, the specific choice of the pulse shapes has a minor effect on the result of the dynamics, as it only determines the minor deviations from adiabatic behavior. Furthermore, the adiabatic evolution is essentially the same for any pulse as long as the following conditions for the control parameters are fulfilled:
\begin{subequations}
\begin{eqnarray}
    \epsilon_j - \epsilon_k \gg v_{jk}\quad t = 0, \\
    \epsilon_k - \epsilon_j \gg v_{jk}\quad t = \tau_\textrm{s},
\end{eqnarray}
\end{subequations}
and that $v_{jk}$ is not zero at the point when $\epsilon_j = \epsilon_k$.

Even though the simulation results depend on 10 input parameters, in the adiabatic case, the frequency of the signal depends only on the ratio of $\Omega$ and $v_{12}$ (see the definition of $\theta_\textrm{so}$ below Eq.~\eqref{eq:nonAB}), which depends on the actual material and device details, and is hard to estimate theoretically. Based on recent shuttling experiments \cite{vanriggelendoelman2023coherent}, we estimate that $v_{12} \sim \Omega$ is very feasible and results in a high frequency in the order of 1 radian per shuttling cycle. In Fig.~\ref{fig:NABP_diabatic}, we have used a conservative choice $\Omega < v_{12}$, leading to oscillations with relatively low frequency.

In Fig.~\ref{fig:NABP_diabatic}, blue, red and yellow data points show simulated results for three different shuttling times (see inset), whereas the solid line shows the adiabatic result of Eq.~\eqref{eq:measurementoutcomeNloop}.
First, the solid line and the blue data points match to a high degree, confirming the analytical result Eq.~\eqref{eq:measurementoutcomeNloop}. 
Second, as the shuttling time is decreased ($\tau_\textrm{s} \lesssim 2$ ns), the mismatch with the adiabatic result grows significantly. We attribute the irregularities developing for shorter shuttling times to diabatic transitions during the shuttling dynamics. 
The critical value of the shuttling time can differ if the shape of the pulses are different. We will turn back to the analysis of the non-adiabaticity in Sec.~\ref{subsec:shuttling_time}.

In Fig.~\ref{fig:NABP_diabatic}b, we illustrate the importance of the relative strength of the Zeeman splitting used to initialise the qubit and thermal fluctuations, focusing on the adiabatic limit for simplicity.
Fig.~\ref{fig:NABP_diabatic}b shows that increasing the Zeeman splitting at a fixed temperature leads to an enhanced contrast of the oscillations, i.e., for a clear illustration of the effect it is beneficial to increase the dc magnetic-field pulse strength and decrease the temperature as much as possible.

The above theory analysis defines an experimental protocol to infer the angle $\theta_\text{so}$ of the single-cycle non-Abelian Berry phase and the  angle $\theta_1$.
For each $N$, the sequence `initialization $\to$ adiabatic cyclic pumping $N$ times $\to$ readout'  should be repeated many times to obtain the probability data points similar to those in Fig.~\ref{fig:NABP_diabatic}.
Those probabilities should then be fitted with a harmonic function of the form \eqref{eq:measurementoutcomeNloop} to extract the angle $\theta_\text{so}$ of the single-cycle non-Abelian Berry phase. 
If the temperature $T$ and the Larmor frequency $\omega_1$ are known, then the only fit parameters are $\theta_\text{so}$ and $\theta_1$. 

We emphasize here, that we have outlined this procedure in terms of a specific minimal model, but the procedure is more general. 
In any model describing a quantum dot loop, the non-Abelian Berry phase can be described as a rotation (of angle $\theta_\text{so}$) on the Bloch-sphere, and the magnetic field used for initialisation and readout defines a Kramers basis to which the axis of the rotation can be compared ($\theta_1$).

\section{Measurement protocol for the internal Zeeman field directions} \label{sec:parametrisation}

\begin{figure}
    \centering
    \includegraphics[width=0.9\linewidth]{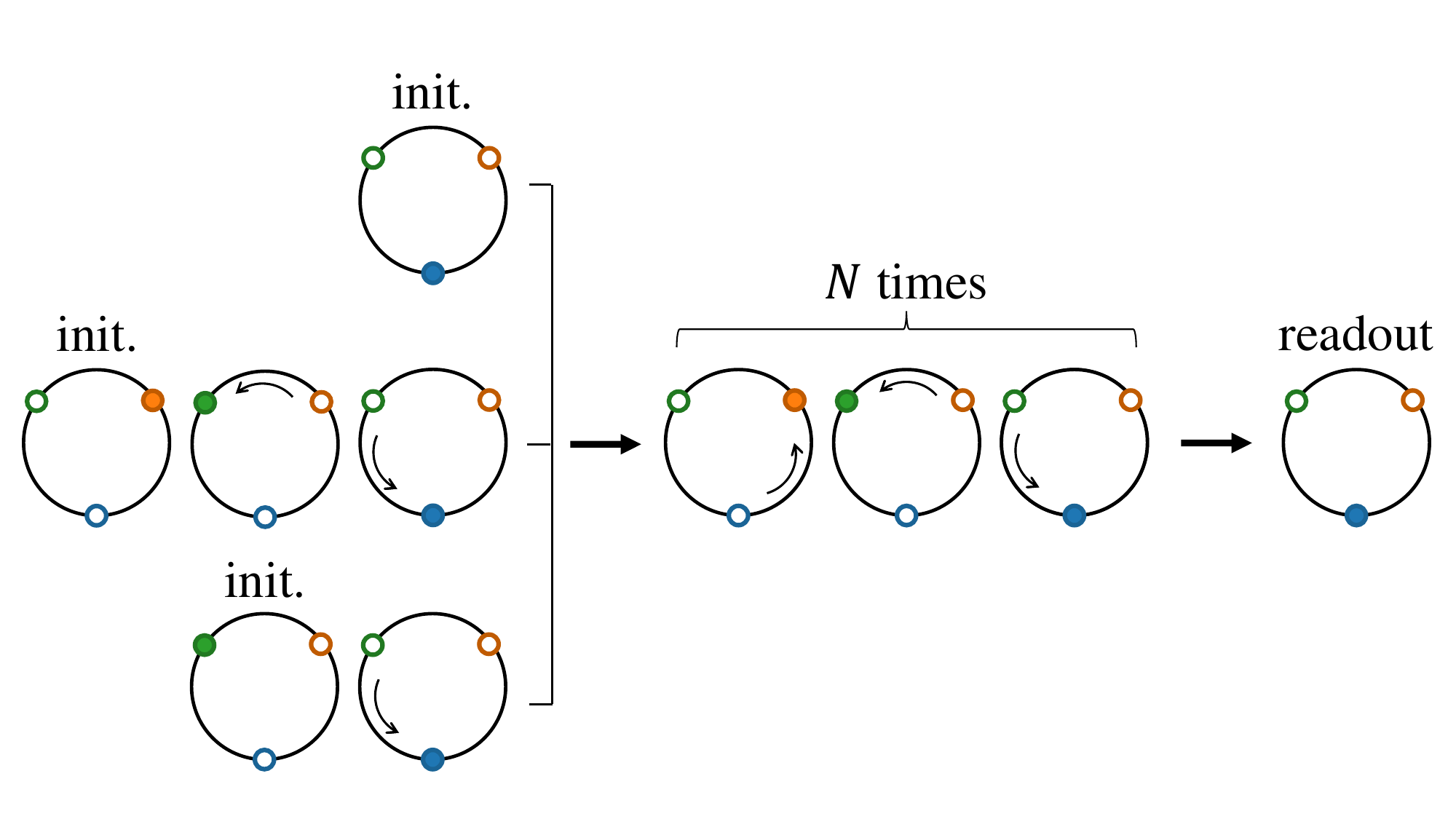}
    \caption{Protocol to infer local Zeeman field directions on the quantum dot of the loop. 
    Three preparation procedures (left) are distinguished. 
    Particle position (filled circle) is shown at the subsequent steps of the protocol. 
    Initialisation (`init.') is done when the particle is localized in one of the dots. 
    Then the particle is shuttled counterclockwise to QD1. 
    After $N$ adiabatic pumping cycles, readout is done in QD1.}
    \label{fig:process}
\end{figure}

A small variation of the previous protocol gives a technique to parametrize the local Zeeman field directions $\theta_2$ and $\phi_2$ in QD2, as well as $\theta_3$ and $\phi_3$ in QD3. 
The setup is the same as before (Fig.~\ref{fig:experiment}a), but this protocol uses two further variants of the initialisation, as shown in Fig.~\ref{fig:process}. 
These variants differ in the location of the particle during the initializing magnetic-field pulse.
After that pulse, the particle is adiabatically pumped, counterclockwise, to QD1. 

Each of these three different preparation procedures results in a different pseudospin state located in QD1, with the pseudospin state depending on the local Zeeman field direction of the QD where the particle was initialized. 
Then, the protocol is continued the same way as in the previous section, i.e., $N$ adiabatic pumping cycles in the loop and then the pseudospin measurement in QD1.
From the measurement data obtained this way, the angles $\theta_2$, $\phi_2$, $\theta_3$, $\phi_3$ describing the local Zeeman field directions in the specific gauge can be inferred. 
This provides a full characterisation of our minimal model described in Sec.~\ref{sec:looppumping_results}. 

Although we have described this protocol using a minimal model, we note that the goal of this protocol and the protocol itself generalises naturally to any model describing adiabatic pumping through a quantum-dot loop. 
In such a generalised model, the angle parameters $\phi_k$ and $\theta_k$ can still be defined as the directions of the local Zeeman fields in the dots in a gauge, where pseudospin is conserved upon tunneling between pairs QD3 $\leftrightarrow$ QD1 and QD2 $\leftrightarrow$ QD3. 

To analyse the protocol, let us fix the same gauge as before in Eq.~\eqref{eqs:conserving_gauge}, i.e., interdot tunneling is pseudospin conserving everywhere except between QD1 and QD2, where it is described by the vector $\boldsymbol{\Omega}$ pointing in the $z$ direction. 
The Zeeman field direction in the $k$-th dot is described with two angles $\theta_k$, $\phi_k$ and we further assume that $\phi_1 = 0$.

In this gauge, the ground state and the excited state can be expressed in the $\ket*{k, \Uparrow}, \ket*{k, \Downarrow}$ basis, when the particle is localized in the $k$-th dot, and the magnetic field is switched on:
\begin{align}
    \ket{k, \text{g}} = \begin{pmatrix}
    \sin\left({\frac{\theta_k}{2}}\right) \\[2pt]
    -e^{i\phi_k}\cos\left(\frac{\theta_k}{2}\right)
    \end{pmatrix}, \ 
    \ket{k, \text{e}} = \begin{pmatrix}
    e^{-i\phi_k}\cos\left({\frac{\theta_k}{2}}\right) \\[2pt]
    \sin\left(\frac{\theta_k}{2}\right)
    \end{pmatrix}.
\end{align}
Thus, after thermalisation, the system is in the following state:
\begin{equation} \label{eq:kdotstate}
    \tilde{\rho}_k = \frac{1}{1 + e^{-\frac{\hbar \omega_{k}}{k_BT}}}  \left(\ket{k,\text{g}}\bra{k,\text{g}} + e^{-\frac{\hbar \omega_{k}}{k_BT}}\ket{k,\text{e}}\bra{k,\text{e}}\right).
\end{equation}
The thermalisation is followed by counterclockwise adiabatic shuttling of the particle to QD1 (e.g., from QD2, on the route $\text{QD2} \to \text{QD3} \to \text{QD1}$), where the tunneling is pseudospin-conserving. 
Therefore, the state $\ket{k, \text{g}}$ ($\ket{k, \text{e}}$) evolves to the state 
\begin{equation}
\ket{\text{g}_k} = 
    \sin \left(\frac{\theta_k}{2} \right) \ket{1, \Uparrow}
    -e^{i\phi_k}\cos \left( \frac{\theta_k}{2} \right)
        \ket{1, \Downarrow},
\end{equation} 
(analogous formula for $\ket{\text{e}_k}$), which has the same amplitudes as $\ket{k, \text{g}}$ in Eq.~\eqref{eq:kdotstate} ($\ket{k, \text{e}}$), but is expressed in the basis $\ket{1, \Uparrow}, \ket{1, \Downarrow}$ localized in QD1.
As a consequence, the state of the system after this preparation procedure reads: 
\begin{equation}
    \rho_k = \frac{1}{1 + e^{-\frac{\hbar \omega_{k}}{k_BT}}}  \left(\ket{\text{g}_k}\bra{\text{g}_k} + e^{-\frac{\hbar \omega_{k}}{k_BT}}\ket{\text{e}_k}\bra{\text{e}_k}\right).
\end{equation}

The next step is to adiabatically pump the particle through the loop $N$ times. 
The probability of the excited state occupation in QD1 after $N$ cycles reads: 
\begin{align} \label{eq:upspinNloop_k}
    P_{k, \textrm{e},N} &= \bra{1, \textrm{e}} U_\text{cycle}^{N} \rho_k U_\text{cycle}^{\dagger N} \ket{1, \textrm{e}} = \nonumber \\ 
    &= a_k - b_k \cos(N\theta_\text{so} - \phi_k), 
\end{align}
where $\theta_\text{so} = 2\arctan\left(\frac{\hbar\Omega}{v_{12}}\right)$, as before, and we introduced the parameters:
\begin{subequations}
\begin{align}
    a_k (\theta_k, \theta_1, T, \omega_{k}) &= \frac{1}{2} - \frac{\tanh\left(\frac{\hbar \omega_{k}}{2k_BT} \right) \cos\theta_1\cos\theta_k }{2},\\
    \label{eq:contrastbk}
    b_k (\theta_k, \theta_1, T, \omega_{k}) &= \frac{\tanh{\left(\frac{\hbar \omega_{k}}{2k_BT}\right)}\sin\theta_1\sin\theta_k}{2}.
\end{align}     
\end{subequations}

\begin{figure}
    \centering
    \includegraphics[width=0.95\linewidth]{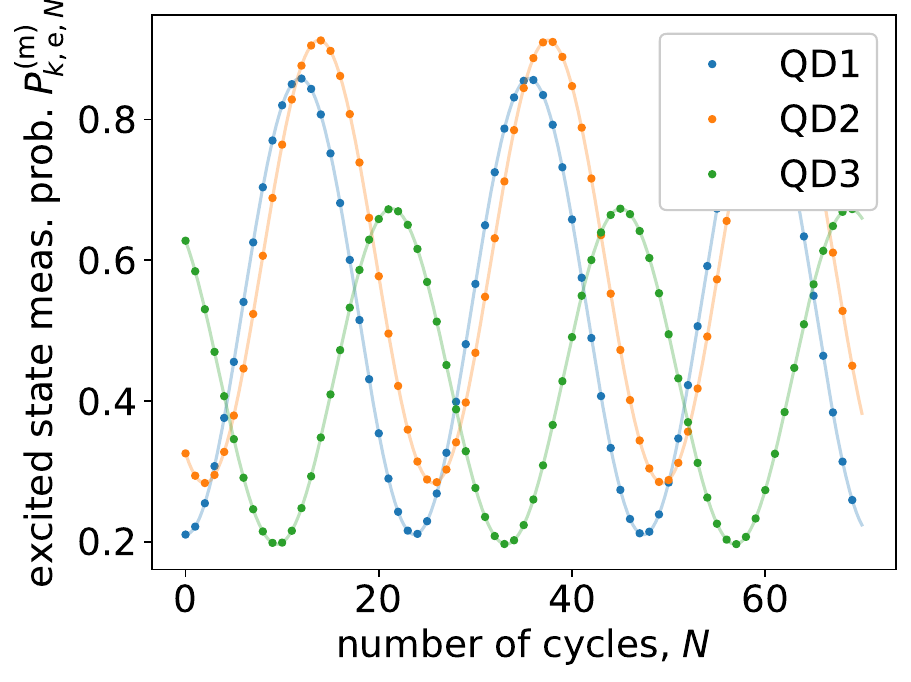}
    \caption{The effect of the non-Abelian Berry phase depending on the qubit position during initialisation. 
    Dots are simulation results, solid lines are analytical results of Eq.~\eqref{eq:measurementoutcomeNloop}. 
    Simulation shows agreement with the analytics.
    Parameters: $\tau_\textrm{s} = 2.5$ ns, $\epsilon_k = 30 \mueV,\, v_{12} = 30 \mueV,\,v_{23} = 28 \mueV,\,v_{31} = 32 \mueV,\,\hbar\Omega = 4\mueV, \, T = 15\,\text{mK},\, B = 10\,\text{mT},\,g_1 = 11,\,g_2 = 10.2,\,g_3 = 15,\,\theta_1 = 72\text{\textdegree},\,\phi_1 = 0\text{\textdegree},\,\theta_2 = 102.9\text{\textdegree},\,\phi_2 = 30\text{\textdegree},\,\theta_3 = 36\text{\textdegree},\,\phi_3 = 144\text{\textdegree}$, where the parameters $\epsilon_k$ and $v_{jk}$ are the amplitudes of the applied sine and sine-squared of sine pulses for the shuttlings shown in Fig.~\ref{fig:experiment}b.}
    \label{fig:parametrisation_sim}
\end{figure}

As a refinement of this result, we consider the same type of readout error as in the previous section.
This implies that the oscillation contrast of the measured excited-state probability $P^{(\text{m})}_{k,\text{e},N}$ is reduced with respect to $b_k$ in Eq.~\eqref{eq:contrastbk} by the same thermal damping factor  $A\left(\frac{\hbar\omega_1}{2k_\textrm{B}T}\right)$ as in Eq.~\eqref{eq:amp}.
We numerically simulated the resulting excited-state probabilities in the limit of adiabatic pumping, and show results in Fig.~\ref{fig:parametrisation_sim}.

Using measurement data obtained via this protocol, one can infer the parameters of the model in Eq.~\eqref{eqs:conserving_gauge}. 
To this end, the measurement data obtained through the three variants (as shown in Fig.~\ref{fig:parametrisation_sim}) is fitted simultaneously, e.g., using $\theta_\text{so}$, $\theta_{1}$, $\theta_2$, $\phi_2$, $\theta_3$, and $\phi_3$ as fit parameters, assuming that temperature $T$ and Larmor frequencies $\omega_1$, $\omega_2$ and $\omega_3$ are known.
We note that $\theta_\text{so}$ is the only joint fit parameter of the three data sets, and it is the common `frequency' characterising the oscillation of all $P^{(\text{m})}_{k, \text{e}, N}$ as functions of $N$.

\section{Discussion} \label{sec:discussion}

In this section, we discuss potential challenges and extensions of our proposed experimental protocol.

\subsection{Optimizing the shuttling time: diabatic transitions, hyperfine interaction and charge noise} \label{subsec:shuttling_time}

We expect that the experimental protocols we have proposed will work if the shuttling time is adjusted properly: It should neither be too fast, nor too slow. 
Too fast shuttling leads to diabatic transitions, leading to errors, such as those exemplified in Fig.~\ref{fig:NABP_diabatic}a.
Too slow shuttling enables qubit decoherence due to various noise types. 

We estimate a lower bound on the shuttling time using the Landau-Zener description for the diabatic transition in a two-level system with Hamiltonian:
\begin{equation}
    H = \begin{pmatrix}
        \frac{\delta \epsilon (t)}{2} & v \\
        v & -\frac{\delta \epsilon (t)}{2}
    \end{pmatrix},
\end{equation}
where $\delta \epsilon (t) = \alpha t$. 
In the Landau-Zener setup, dynamics starts in the ground state at $t = -\infty$ and ends at $t = \infty$. 
The diabatic transition probability ($P_\textrm{d}$) is given by the Landau-Zener formula \cite{Zener1932Non-adiabaticLevels}:
\begin{equation}
    P_\textrm{d} = e^{-\frac{v^2}{h \alpha}},
\end{equation}
To describe the dynamics in our shuttling setup, we estimate the sweep velocity as $\alpha = \frac{\epsilon \pi}{\tau_\textrm{s}}$. 
Substituting the realistic parameter values $\epsilon = 20v = 1.2$ meV, for a $P_\textrm{d} < 10^{-4}$ suppression of the diabatic transition, we obtain the following constraint for the shuttling time:
\begin{equation}
    \tau_\textrm{s} \gtrsim 4 \, \textrm{ns}.
\end{equation}
This Landau-Zener description applies to a different type of pulse sequence than the one we used in our simulation. In that case, a more sophisticated method is required to obtain the analytic lower bound of the shuttling time which is omitted here. However, the simulation results in Fig.~\ref{fig:NABP_diabatic}a and the above estimation support the fact that the adiabatic shuttling of the particle between two dots is achievable within few nanoseconds.

We estimate an upper bound on the shuttling time by considering the effect of hyperfine interaction, caused by nuclear spins in the semiconductor hosting the quantum dots.
Hyperfine interaction is an important and well-described cause of dephasing in semiconductor spin qubits. 
Using a specific dephasing model \cite{Fischer2008SpinDot}, the qubit inhomogeneous dephasing time $T_2^*$ caused by hyperfine interaction is expressed as:
\begin{equation} \label{eq:hyperfine_dephasing}
    \frac{\hbar^2}{T_2^{*2}} = \frac{1}{2N_n} \nu I(I+1) A_n^2, 
\end{equation}
where
$\nu$ is the abundance of the isotope having nuclear spin $I$,
$N_n = \pi a^{*2}_\textrm{B}w/v_0$ is the number of atoms in the quantum dot
($a^{*}_\textrm{B}$ is the effective Bohr radius of the dot, $w$ is the width of the quantum well, $v_0$ is the atomic volume of the host material),
and $A_n$ is the hyperfine interaction strength.

We estimate $T_2^* \approx 1.4\, \mu$s from Eq.~\eqref{eq:hyperfine_dephasing}, using realistic parameter values for holes in natural germanium \cite{Philippopoulos2020HyperfineNanostructures}:
$a^{*}_\textrm{B} = 50$ nm, $w = 15$ nm, $v_0 = 2.26 \cdot 10^{-2} \,\textrm{nm}^3$, $\nu = 0.0776$, $I = 9/2$, and $A_n = -1.1$ $\mueV$. 
The dephasing time gives a condition for the shuttling time and the number of cycles which reads as
\begin{equation}
    3N\tau_\text{s} \lesssim T_2^{*}.
\end{equation}
With the substitution of $\tau_s = 4$ ns, we get an upper bound for the number of coherent cycles: $N \lesssim 100$.
Hence, we conclude that our proposed experiment is feasible with hole spin qubits in natural germanium.
Motional narrowing \cite{Flentje2017CoherentSpins, Mortemousque2021EnhancedDots} and isotopic purification can further suppress effects of hyperfine interaction and hence increase this feasibility. 

Charge noise, including fluctuating charge traps, gate-referred noise, and electron-phonon interaction, could also affect the feasibility of the experiment \cite{Krzywda2020AdiabaticNoise, Krzywda2021InterplayDots}. 
Based on experimental results on the coherence time of germanium hole spin qubits \cite{Hendrickx2021AProcessor, vanriggelendoelman2023coherent, hendrickx2023sweetspot}, we conclude that the effect of charge noise for our proposed experiments is probably insignificant. Our argument is as follows. Static qubits were investigated in Ref.~\cite{Hendrickx2021AProcessor} with Larmor frequencies in the range of 2-4 GHz, and the observed coherence times $T_2^{*}$ were 150-400 ns. In Ref.~\cite{vanriggelendoelman2023coherent} the same device was used to demonstrate coherent shuttling. The Larmor frequencies of the qubits were between 0.8-1.1 GHz, and characteristic decay constants (number of coherent shuttlings) $n^{*} =64-77$ was reported, which maps to $T^{*}_2 = 320-385$ ns with shuttling time $\tau_\textrm{s} = 4$ ns and 1 ns of idle time between each shuttling. In Ref.~\cite{hendrickx2023sweetspot}, a static qubit with Larmor frequency $f_L = 21$ MHz was measured and the coherence time was found to be $T_2^{*} = 17.6\,\mu$s. These experimental results suggest that effect of the charge noise can be suppressed by decreasing the Larmor frequency as it was predicted theoretically \cite{San-Jose2006GeometricalDots}. Therefore, in our experimental protocol where the Larmor frequency is zero, we expect that the coherence time of few tens of microseconds is achievable which translates to 1000 coherent shuttling cycles.

Charge noise can also induce direct dephasing in zero magnetic field, called geometric dephasing \cite{San-Jose2006GeometricalDots}. 
The geometric dephasing time was estimated as $T_\textrm{geom} \approx 20$ ms for GaAs quantum dots at $T = 50$ mK in Ref.~\cite{San-Jose2006GeometricalDots}. This suggests, that in our protocol the geometrical dephasing is negligible as compared to hyperfine noise. 

To conclude, the time-scale estimates collected in this section suggest the feasibility of the experiments proposed in Sec.~\ref{sec:NABP_meas} and \ref{sec:parametrisation}.

\subsection{Risks and challenges}

\emph{Earth's magnetic field.}
The Earth's magnetic field amounts to a few tens of microteslas.
This may be regarded `weak', but it is strong enough to affect the experiments proposed above. 
In fact, the Larmor precession time corresponding to a magnetic field of $B_E = 65 \, \mu$T with $g$-factor $g = 2$ is $T_L \approx 550\,$ns. 
This Larmor time can be further shortened in materials with strong spin-orbit coupling, e.g., for planar hole quantum dots, where the $g$-factor is enhanced in an out-of-plane magnetic field.
\cite{Winkler2003SpinOrbitSystems, Jirovec2021AGe, hendrickx2023sweetspot}. 
For shuttling times comparable to $T_L$, the non-Abelian adiabatic dynamics would be overwhelmed by the Larmor precession due to the magnetic field. 
The effect can be reduced by using shorter shuttling times, e.g., in the few-nanosecond ballpark discussed above.
To further mitigate this effect, the Earth's magnetic field should be shielded in the experiment, which can be done using well-established methods \cite{Krieger2014DesignLaboratory, Kreikebaum2016OptimizationResonators, Flanigan2016MagneticDetectors, Gordon2022EnvironmentalQubits}.

\emph{Coincidental alignment of the two axes ($\theta_1 = 0$).}
The key feature of the measurement results of the experimental protocol proposed in Sec.~\ref{sec:NABP_meas} is the oscillatory behavior of $P^{(\text{m})}_{1,\text{e},N}$ as the number $N$ of cycles is increased, as shown in Eq.~\eqref{eq:measurementoutcomeNloop} and Fig.~\ref{fig:NABP_diabatic}. 
From Eq.~\eqref{eq:amp}, it is clear that this oscillatory behavior is absent if $\theta_1 = 0$.
The success of the experiment proposed in Sec.~\ref{sec:NABP_meas} relies on the assumption that the Kramers basis $\ket{1, \text{g}}$, $\ket{1,\text{e}}$, defined by the local magnetic field used for initalization and readout, does \emph{not} coincide with the Kramers eigenbasis $\ket{1,\Uparrow}$, $\ket{1,\Downarrow}$ of the non-Abelian Berry phase propagator.
This is formalized as the assumption $\theta_1 \neq 0$.
More generally, the relative angles $\theta_k$ play an important role in determining the contrast of the oscillations described by Eqs.~\eqref{eq:amp} and \eqref{eq:contrastbk}.
In an experiment, these angles, which are determined by the microscopic details of spin-orbit interaction, confinement and potentials, strain patterns, and disorder, can presumably be changed by continuous tuning of gate voltages, or by changing the charge occupations of the dots \cite{Han2023VariableSensing, Crippa2018ElectricalQubits, hendrickx2023sweetspot, abadillouriel2022hole, Martinez2022HoleFields}. 

\emph{Efficient initalization and readout.}
As illustrated in Fig.~\ref{fig:NABP_diabatic}(b), a strong magnetic field pulse and a low temperature are essential for high-fidelity initialization and readout, and consequently, for obtaining high-contrast excited-state probability oscillations in the experiment. 
A current pulse through an on-chip wire in the close vicinity of the quantum dot loop could deliver a millitesla-scale magnetic field \cite{Koppens2006DrivenDot}, although its engineering and fabrication might be challenging. 
Therefore, to improve the visibility of the results one might need to cool the device below 15 mK (i.e., the value we have used in our simulations). 
Note that that the sub-millikelvin cooling of mesoscopic electronic devices is an active research area with recent successes \cite{Huang2007Disappearance30mK, Maradan2014GaAsSensing, Sarsby2020500Nanoelectronics, Samani2022MicrokelvinThermometer}.
Lower temperature has the additional advantage that it reduces the thermal excitation in the shuttling processes. 
One should also be careful about switching off the magnetic field instantaneously, because it can lead to leakage to higher level orbital states. The leakage is proportional to $g^* \mu_\textrm{B} B/\Delta$ in first order, where $g^* \mu_\textrm{B} B$ is the characteristic energy scale of the Zeeman splitting, and $\Delta$ is the gap between the ground-state orbital and the first excited orbital of the confinement. This is in the order of $10^{-3}$ for $B=5$ mT with a typical orbital level spacing of $\Delta = 1$ meV. Thus, the leakage can be neglected even if the magnetic field is switched off instantaneously, let alone in the case of a smooth switch-off which is realistic in an experiment.

\subsection{An alternative protocol without magnetic-field pulses: using the singlet state for initialisation and readout}\label{sec:singletinit}

\begin{figure}
    \centering
    \includegraphics[width=0.95\linewidth]{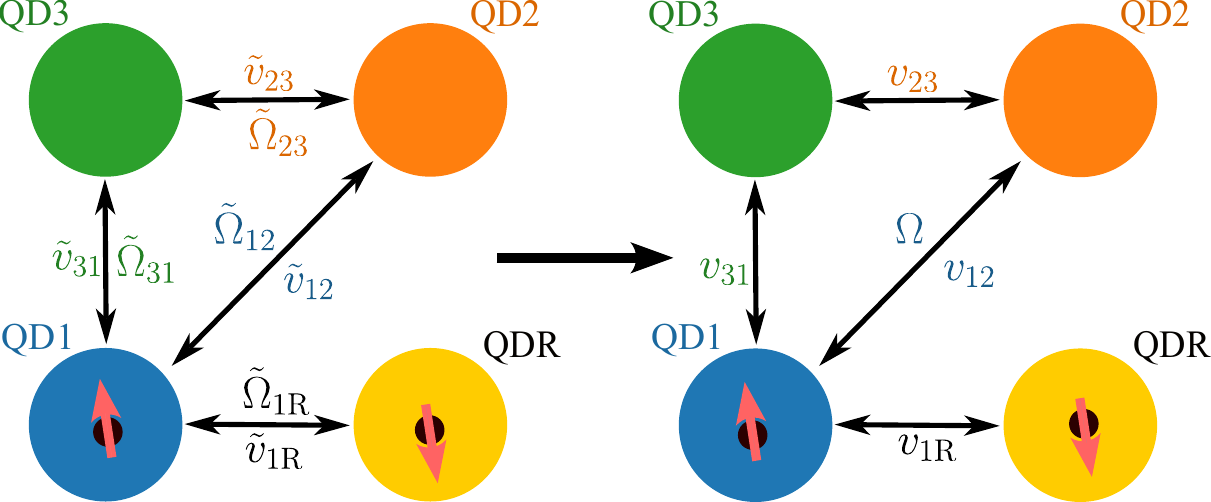}
    \caption{Alternative protocol to measure the non-Abelian Berry phase using a reference dot and the Pauli blockade mechanism for initialization and readout. 
    The advantage of this protocol is that direct dot--reservoir tunneling and magnetic-field pulses are not required.
    Initialization is done by separating a two-qubit singlet state in QDR into QDR and QD1; 
    readout is done after $N$ adiabatic pumping cycles QD1 $\to$ QD2 $\to$ QD3 $\to$ QD1, by spin-to-charge conversion using Pauli blockade between QD1 and QDR. 
    Left panel illustrates the tunneling terms of the Hamiltonian in an arbitrary gauge; right panel does the same in a convenient gauge where all tunneling terms are pseudospin-conserving, except the one on the QD1 $\leftrightarrow$ QD2 bond.}
    \label{fig:fourdot_array}
\end{figure}

In the previous section, we have discussed the challenges of creating a strong-enough magnetic-field pulse and a low-enough temperature for our experimental protocols. 
As a possible workaround of these, we now provide  an alternative protocol, which requires an extra reference dot, denoted as QDR in Fig.~\ref{fig:fourdot_array}, and utilizes a two-qubit singlet state in that reference dot for initialization and readout, which is a well-known procedure for initialization \cite{Petta2005AppliedDots}. 
This protocol can reveal the angle $\theta_\text{so}$ characterizing the non-Abelian Berry phase, without requiring a magnetic-field pulse. 
Furthermore, the low-temperature requirement is also relaxed, because initialization and readout is based on Pauli spin blockade, where the energy scale (the two-qubit singlet-triplet splitting in QDR) competing with temperature can be made much greater than the Zeeman splitting in a few millitesla \cite{Petta2005AppliedDots}.

The layout for this alternative protocol is shown in Fig.~\ref{fig:fourdot_array}. 
To simplify the discussion, we use a specific gauge (i.e., specification of the Kramers-pair basis in the dots) where tunneling is pseudospin-conserving on the bonds 
QDR $\leftrightarrow$ QD1, 
QD2 $\leftrightarrow$ QD3,
and 
QD3 $\leftrightarrow$ QD1,
and it is pseudospin-non-conserving only on the bond
QD1 $\leftrightarrow$ QD2.
In this protocol, initialization consists of thermalizing a two-qubit singlet state in QDR, and separating the two particles with an adiabatic shuttling step, by tuning the detuning and tunneling between QDR and QD1 in time.
Before this shuttling, the double quantum dot formed by QDR and QD1 is in the (2,0) charge configuration regime and its pseudospin degree of freedom is in the singlet state. During the shuttling, the parameters are tuned adiabatically such that the system prefers the (1,1) charge configuration. Thus, after the shuttling, the double quantum dot is in the (1,1) charge configuration occupied by a pseudospin-singlet state.
Ideally, the tunneling matrix element $v_{1\text{R}}$ is switched off exactly before and after the shuttling step. 
This initialization procedure results in the separated singlet configuration depicted in Fig.~\ref{fig:fourdot_array}.
 
During a single pumping cycle QD1 $\to$ QD2 $\to$ QD3 $\to$ QD1, the particle in QD1 is pumped around.
Therefore, the time evolution of the two-qubit system is governed by the following unitary, expressed in the two-qubit basis $\ket{\Uparrow\Uparrow}, \ket{\Uparrow\Downarrow}, \ket{\Downarrow\Uparrow}, \ket{\Downarrow\Downarrow}$, with the first (second) arrow referring to QD1 (QDR):
\begin{equation}
    V = \begin{pmatrix}
        e^{i\theta_\text{so}} & 0 & 0 & 0 \\
        0 & e^{i\theta_\text{so}} & 0 & 0 \\
        0 & 0 & e^{-i\theta_\text{so}} & 0 \\
        0 & 0 & 0 & e^{-i\theta_\text{so}}
    \end{pmatrix}.
\end{equation}
Since the initial two-qubit state is the singlet state
\begin{equation}
\ket{S} = \frac{1}{\sqrt{2}}\left(\ket{\Uparrow\Downarrow} - \ket{\Downarrow\Uparrow}\right),
\end{equation}
after $N$ pumping cycles, it evolves into
\begin{align}
\label{eq:evolvedsinglet}
V^N\ket{S} &= \frac{1}{\sqrt{2}}\left(e^{iN\theta_\text{so}}\ket{\Uparrow\Downarrow} - e^{-iN\theta_\text{so}}\ket{\Downarrow\Uparrow}\right) = \nonumber \\ 
&= \cos(N\theta_\text{so})\ket{S} + i\sin(N\theta_\text{so})\ket{T_0}.
\end{align} 

At this point, the QDR and the single particle in it can be utilized once again, now for readout via spin-to-charge conversion based on Pauli blockade. 
Specifically, an attempt to shuttle the particle in QD1 to QDR is made, and charge sensing is done on the dots to reveal if the final state of QDR contains one or two particles. 
Due to Pauli blockade, one (two) particle(s) measured in QDR implies that the state $V^N \ket{S}$ was triplet (singlet). 
The singlet measurement probability is expressed from Eq.~\eqref{eq:evolvedsinglet} as
\begin{equation}
    P_S = |\bra{S}V^N\ket{S}|^2 = \cos^2(N\theta_\text{so}).
    \label{eq:oscillations}
\end{equation}
The measurement data obtained for this singlet probability, as a function of the number $N$ of the cycles, should show oscillations, whose `frequency' reveals the angle of the non-Abelian Berry phase. 

This protocol has advantages over the one described in Sec.~\ref{sec:NABP_meas}, which uses magnetic-field-enabled initialisation and readout.
First, no additional wire is required to create the magnetic-field pulse. 
Devices without that element are readily available in the $2\times 2$ quantum dot array layout\cite{Hendrickx2021AProcessor, borsoi2022shared, hendrickx2023sweetspot, vanRiggelen2022PhaseQubits, wang2022probing, vanriggelendoelman2023coherent}. 
Second, the oscillation of the measured singlet probability is  not suppressed by the factor of $\sin^2\theta_1$ (i.e., the angle between the axis of the non-Abelian Berry phase and the local Zeeman axis) appearing in Eq.~\eqref{eq:amp}.
Third, the effect of the finite temperature, e.g., appearing in the suppression factor $A(x)$ in Eq.~\eqref{eq:amp}, is weakened, as the singlet-triplet splitting in QDR can be made much greater than the Zeeman splitting caused by a few millitesla magnetic field.

A challenging component of this alternative protocol may be to sufficiently reduce the exchange interaction $J$ between the two spins, after they have been separated at the end of the initialization step. 
A nonzero exchange interaction strength introduces a nonzero singlet-triplet energy splitting in the two-particle spectrum, and hence a corresponding timescale $h/J$, which poses an upper limit on the timescale $3 N \tau_\textrm{s}$ of the $N$ pumping cycles. 
That is, the condition $\frac{3N\tau_\textrm{s}J}{h} \ll 1$ should be satisfied to neglect the effect of the residual exchange interaction. 
Substituting $\tau_\textrm{s} = 3$ ns and $N = 50$, we obtain the condition  $J/h \ll 2.3$ MHz. 

Also, if the exchange interaction is nonzero, charge noise can in principle corrupt the two-qubit state. However, this is probably not the a critical threat for the proposed alternative protocol, for the following reasons. In the initialization step, when the exchange interaction is strong between the pseudospins, the system is in its ground state, and the evolution is adiabatic, hence charge noise has no effect on it as long as it is quasi-static and not resonant with the singlet-triplet splitting. Experimental results confirm this as well, see, e.g.~Ref.~\cite{Petta2005AppliedDots}. After the separation of the particles, the exchange interaction is turned off completely, at least in an ideal scenario. The remnant non-zero exchange interaction is a potential threat to the alternative protocol during the circular shuttling cycles, but its quasi-static fluctuation is irrelevant as long as the condition $J/h \ll 2.3$ MHz is met during the process.

An additional challenge one might think of is that strong spin-orbit coupling can make Pauli blockade readout dysfunctional, as reported in many papers, e.g.~in Ref.~\cite{Hendrickx2021AProcessor}. However, this is not a problem in this protocol because of the absence of an external magnetic field. In this case, in a well chosen gauge, in which the tunneling between QDR and QD1 is pseudospin-conserving, the pseudospin is a conserved quantity, and therefore Pauli blockade is not lifted and can be utilized for readout.

\subsection{Setups for proving the non-Abelian nature of the geometric quantum gate}

\begin{figure}
    \centering
    \includegraphics[width=\linewidth]{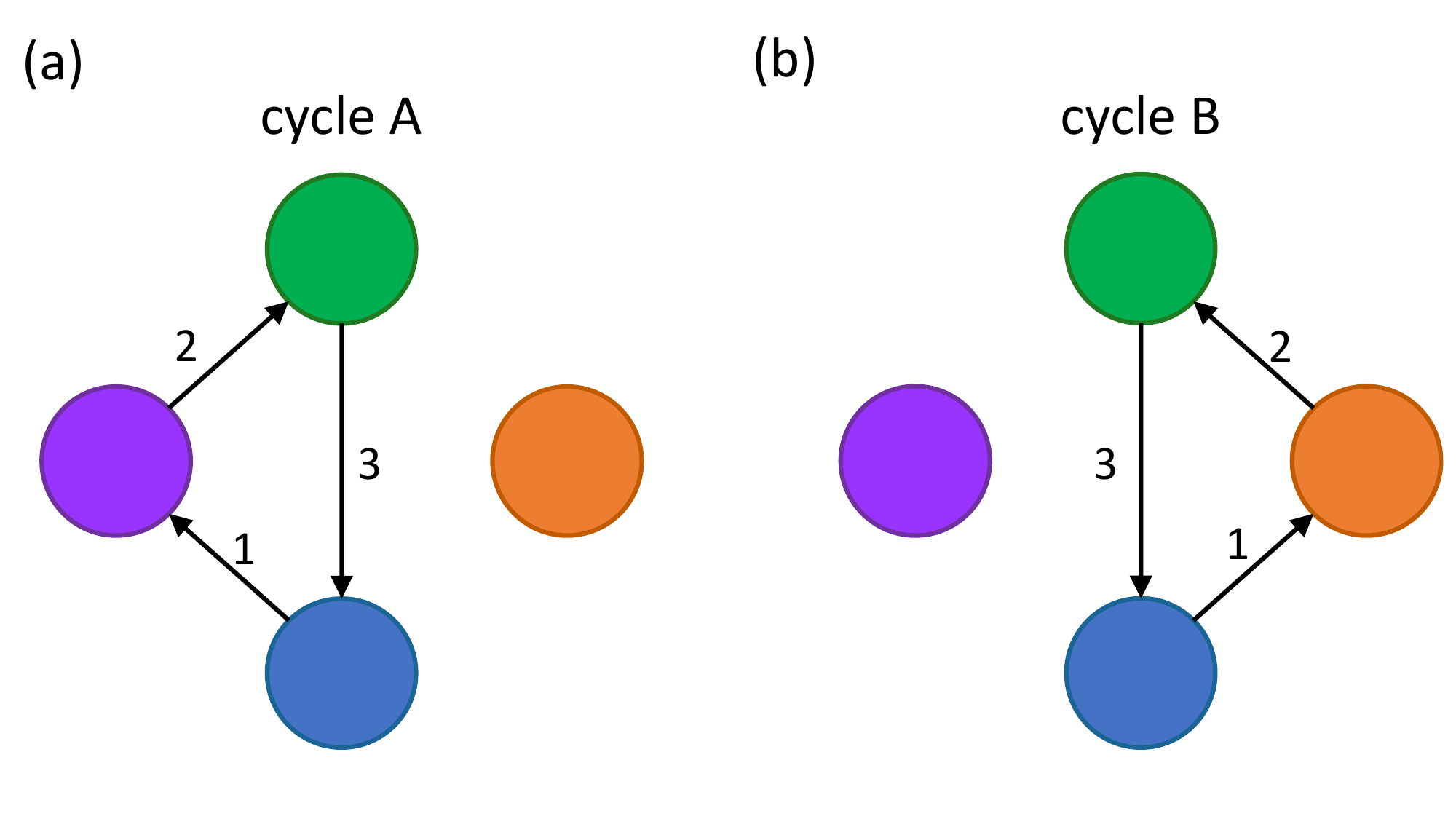}
    \caption{Revealing the non-Abelian character of the geometric quantum gates. Panels (a) and (b) depict two different pumping cycles A and B in a $2\times 2$ quantum dot array. 
    In general, performing cycle A first and cycle B second leads to a different measurement result then shuttling in the reverse order. 
    }
    \label{fig:NA_measurement}
\end{figure}

A possible generalisation of the proposed measurement is to show that the induced geometric quantum gate in a quantum dot loop is truly non-Abelian. 
One way to achieve this is by extending the loop in Fig.~\ref{fig:experiment}(a) with an additional quantum dot. Then the qubit can be driven through two different cycles after each other, as shown in Fig.~\ref{fig:NA_measurement}(a) and (b). By showing that the induced transformation depends on the ordering of the different cycles, one can conclude that the geometric single-qubit gates induced by cycle A or cycle B do not commute.

The demonstration of the non-Abelian character of the geometric gate could also be attempted using three quantum dots only, by utilising nearby orbitals in QD2 and QD3. 
To discuss a simple example, we denote the charge configuration of the three dots as $(N_1,N_2,N_3)$, where $N_j$ denotes the number of particles in QD $j$, counted with respect to a certain filled-shell configuration.
Using this notation, the cycle QD1 $\to$ QD2 $\to$ QD3 $\to$ QD1 studied so far corresponds to, e.g., 
$(1,0,0) \to (0,1,0) \to (0,0,1) \to (1,0,0)$; call this cycle A.
A different cycle, say, cycle B, can be defined using the same three quantum dots by shifting the charge configuration of QD2 and QD3 such that the cycle takes the form
$(1,2,-2) \to (0,3,-2) \to (0,2,-1) \to (1,2,-2)$.
Since spin-orbit features, e.g., spin-dependent tunneling, is expected to vary between different shells \cite{Han2023VariableSensing},
hence the non-Abelian Berry phase corresponding to cycle A and B are expected to be different, and in general, non-commuting. 

\subsection{Relation of our proposals to a recent `triangular shuttling' experiment}

In a recent experiment \cite{vanriggelendoelman2023coherent}, coherent shuttling of a spin qubit through a quantum dot loop was demonstrated. 
In that manuscript, the `quantization axis tilt angles' of tunnel-coupled quantum dot pairs were determined by spectroscopy, and also by using time-resolved shuttling experiments.
Both experiments were carried out in a nonzero magnetic field with a fixed direction. 

\begin{figure}
    \centering
    \includegraphics[width=\linewidth]{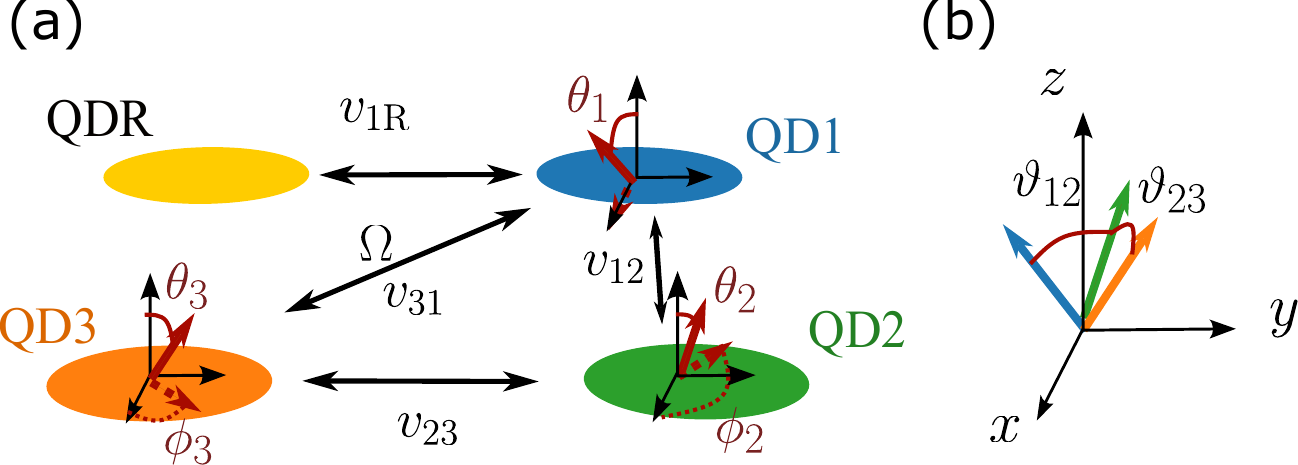}
    \caption{Relation of the local Zeeman field directions and the quantization axis tilt angles. 
    (a) Layout of the device used in Ref.~\cite{vanriggelendoelman2023coherent}.
    Red solid vectors depict local Zeeman field directions $\mathbf{n}_1$,
    $\mathbf{n}_2$,
    $\mathbf{n}_3$.
    Labels on the tunnel couplings indicate the gauge choice: pseudospin-non-conserving tunneling ($\Omega$) appears only on the bond between QD2 and QD3. 
    (b) Quantization axis tilt angles  $\vartheta_{12}$ and $\vartheta_{23}$ that were measured in Ref.~\cite{vanriggelendoelman2023coherent}. 
    Blue/green/orange arrow shows the local Zeeman field direction in QD1/QD2/QD3.}
    \label{fig:delft}
\end{figure}

Here, we establish the relation between the quantization axis tilt angles and the parameters in our model. 
In Fig.~\ref{fig:delft}(a), we draw the four-dot configuration used in \cite{vanriggelendoelman2023coherent}.
In Fig.~\ref{fig:delft}(a), we also indicate the gauge choice that is used in \cite{vanriggelendoelman2023coherent}: tunnel couplings between 
QDR $\leftrightarrow$ QD1,  
QD1 $\leftrightarrow$ QD2, and 
QD2 $\leftrightarrow$ QD3 are transformed to be pseudospin-conserving, and the tunnel coupling between 
QD3 $\leftrightarrow$ QD1
is retained as pseudospin-non-conserving.
In this setting, the system of the three numbered QDs is described by the Hamiltonian in Eq.~\eqref{eqs:conserving_gauge}, with the only difference that in $H_\text{stun}$, the operator $\tau^y_{12}$ is replaced by $\tau^y_{23}$.

In this gauge, at a finite magnetic field, the quantization axes of the numbered dots are given by the unit vectors $\mathbf{n}_k$, see below Eq.~\eqref{eqs:conserving_gauge}.
We illustrate these quantization axes in Fig.~\ref{fig:delft}(a), where the thick red arrows depict the vectors $\mathbf{n}_k$ (arbitrary example), and the corresponding polar and azimuthal angles $\theta_k$ and $\phi_k$ are also indicated.
The quantization axes $\mathbf{n}_k$ are shown together in Fig.~\ref{fig:delft}(b), where the angle enclosed by $\mathbf{n}_1$ and 
$\mathbf{n}_2$ is denoted by $\vartheta_{12}$, 
and the angle enclosed by 
$\mathbf{n}_2$ and 
$\mathbf{n}_3$ is denoted by $\vartheta_{23}$.
These angles $\vartheta_{12}$ and $\vartheta_{23}$ were determined in the experiment \cite{vanriggelendoelman2023coherent}. 
The relation between these quantization axis tilt angles and the angle parameters of our Hamiltonian
reads:
\begin{equation}
    \vartheta_{jk} = \arccos \left( \cos (\phi_j - \phi_k)\sin \theta_j \sin \theta_k + \cos\theta_j \cos\theta_k \right),
    \label{eq:tilt}
\end{equation}
where $jk \in \{12,23\}$.

Note that the protocol we suggest in 
Sec.~\ref{sec:parametrisation} enables a more complete characterization of the Hamiltonian than the inference of the tilt angles; i.e., our protocol enables the inference of all five angle parameters $\theta_1$, $\theta_2$, $\phi_2$, $\theta_3$, $\phi_3$ 
(from these, the quantization axis tilt angles can be calculated using Eq.~\eqref{eq:tilt}), as well as the strength $\Omega$ of the pseudospin-non-conserving tunneling term.

The device and protocols used in Ref.~\cite{vanriggelendoelman2023coherent} are very similar to those described in our present work, highlighting the near-term feasibility of our proposals.
The device of \cite{vanriggelendoelman2023coherent} has no on-chip pulsed magnetic field source, and utilizes Pauli blockade for initialisation and readout.
As a consequence, it enables the direct implementation of our alternative protocol described in Sec.~\ref{sec:singletinit}, and correspondingly, the inference of the angle of rotation of the holonomic quantum gate induced on the qubit by a single pumping cycle through the quantum dot loop. 
In fact, the `triangular shuttling' protocol, which is carried out in a nonzero static magnetic field in Ref.~\cite{vanriggelendoelman2023coherent}, is identical to our alternative protocol (Sec.~\ref{sec:singletinit}), if the magnetic field is switched off in the experiment. 

Based on Ref.~\cite{vanriggelendoelman2023coherent}, we identify the slow diagonal shuttling (shuttling time of $\tau_\textrm{s} \approx 36$ ns) as a potential bottleneck to observe the oscillations predicted by Eq.~\eqref{eq:oscillations}. 
As we described above, the Earth's magnetic field and hyperfine interaction set an upper bound on the time window of the experiment (presumably around one microsecond), and the qubit dynamics induced by the non-Abelian Berry phase may be masked by those magnetic effects if shuttling is too slow. 
Improvements are expected if the diagonal shuttling is sped up (e.g., by using a dedicated barrier gate addressing the corresponding tunnel barrier), by shielding the Earth's magnetic field, and by using isotopically purified materials to suppress the hyperfine effect.

\section{Conclusion} \label{sec:conclusion}

We provided a theoretical description of the non-Abelian Berry phase or geometric quantum gate of a spin qubit, induced upon an adiabatic shuttling cycle through a quantum-dot loop in the presence of spin-orbit interaction and absence of magnetic field. 
Time-reversal symmetry is expected to provide an extra level of protection of the Kramers degeneracy exploited in this setup, in contrast to earlier experiments where the degeneracy relies on fine-tuning of system parameters. 
We highlighted two experimental protocols to detect features of the non-Abelian Berry phase: the first (second) one relies on a pulsed magnetic field (zero-field Pauli spin blockade) for initialisation and readout.
Furthermore, we predict that a variant of the first protocol can be used to infer the parameters of the spin-orbit-coupled Hamiltonian.
We expect a near-term realisation of our protocol, as all key elements of it have been demonstrated in spin-qubit experiments. 
Such a realisation would be important to assess the potential of holonomic quantum gates for spin-based quantum information processing. 

\begin{acknowledgments}
We thank G.~F\"ul\"op, A.~Geresdi, F. van Riggelen-Doelman, L.~ Vandersypen, and M.~Veldhorst for useful discussions. 
This research was supported by the Ministry of Culture and Innovation and the National Research, Development and Innovation Office (NKFIH) within the Quantum Information National Laboratory of Hungary (Grant No.~2022-2.1.1-NL-2022-00004), by NKFIH via the OTKA Grant No.~132146, and by the European Union's Horizon Europe research and innovation programme via the IGNITE project.
\end{acknowledgments}

\appendix
\section{Gauge transformation in the three dot loop}
\label{app:gauge}

In this appendix, we formulate the gauge transformation shown in Fig.~\ref{fig:dotloop_gauge}, i.e.~the transformation of the tunneling terms of the Hamiltonian from the form of Eq.~\eqref{eq:hamiltonian} to Eq.~\eqref{eq:tunaftergauge}.
The tunneling terms of the Hamiltonian \eqref{eq:tildetun} and \eqref{eq:tildestun} in the block form expanded in the orbital degree of freedom reads as:
\begin{equation}
\label{eq:H3by3}
    H_\textrm{tun} + H_\textrm{stun} = \begin{pmatrix} 0 & \Tilde{v}_{12} - i \hbar\mathbf{\Tilde{\Omega}}_{12} \cdot \boldsymbol{\tilde \sigma} & \Tilde{v}_{31} + i \hbar\mathbf{\Tilde{\Omega}}_{31} \cdot \boldsymbol{\tilde \sigma} \\
    h.c.  & 0 & \Tilde{v}_{23} - i \hbar\mathbf{\Tilde{\Omega}}_{23} \cdot \boldsymbol{\tilde\sigma} \\ 
    h.c. & h.c. & 0\end{pmatrix},
\end{equation}
where $\boldsymbol{\tilde \sigma}$ are the Pauli matrices defined with the basis $\left(\ket*{\tilde{\Uparrow}},\ket*{\tilde{\Downarrow}}\right)$.  Let us first choose an appropriate basis in QD3 to eliminate the pseudospin-non-conserving tunneling between QD1 and QD3, that is, to eliminate $\mathbf{\Tilde{\Omega}}_{31}$ in Eq.~\eqref{eq:H3by3}. 
We do this by choosing new local Kramers pairs to define the local Pauli operators; these new local Kramers pairs $\ket*{\hat{\Uparrow}},\ket*{\hat{\Downarrow}}$ are chosen as
\begin{equation}
    \ket*{k,\hat \Downarrow} = U_3\ket*{k, \tilde \Downarrow} , \ket*{k,\hat \Uparrow} = U_3\ket*{k, \tilde \Uparrow},
\end{equation}
where $U_3$ is a local pseudospin rotation in QD3 with rotation axis $\boldsymbol{\tilde{\omega}}_{31} = \boldsymbol{\Tilde{\Omega}}_{31}/|\boldsymbol{\tilde{\Omega}}_{31}|$ and angle $\Tilde{\theta}_\text{so}^{(31)} = 2\arctan\frac{\hbar|\boldsymbol{\Tilde{\Omega}}_{31}|}{\Tilde{v}_{31}}$ in the clockwise (negative) direction. Explicitly, it reads:
\begin{equation} \label{eq:U3}
    U_3 = e^{-i\frac{\Tilde{\theta}_\text{so}^{(31)}}{2} \boldsymbol{\Tilde{\omega}}_{31} \cdot \tilde{\boldsymbol{\sigma}} \otimes \tau_3}.
\end{equation}
Note that the Kramers basis states on QD1 and QD2 are unchanged by $U_3$.

The Pauli operators in the new gauge, denoted as $\boldsymbol{\hat \sigma}$, are related to the original Pauli operators $\boldsymbol{\tilde \sigma}$ as
\begin{equation}
    \boldsymbol{\hat \sigma} 
    \otimes \tau_k= U_3 
    \left(
    \boldsymbol{\tilde \sigma} 
    \otimes \tau_k 
    \right)
    U_3^\dag,
\end{equation}
for all $k \in \{1,2,3\}$.
Note that the unitary $U_3$ has the same form in the two gauges, i.e.~$U_3 = e^{-i\frac{\Tilde{\theta}_\text{so}^{(31)}}{2} \boldsymbol{\Tilde{\omega}}_{31} \cdot \tilde{\boldsymbol{\sigma}} \otimes \tau_3} = e^{-i\frac{\Tilde{\theta}_\text{so}^{(31)}}{2} \boldsymbol{\tilde{\omega}}_{31} \cdot \boldsymbol{\hat \sigma} \otimes \tau_3}$. 
\onecolumngrid
Using this identity, the tunneling Hamiltonian in the new gauge reads as:
\begin{align}
    H&_\textrm{tun} + H_\textrm{stun} = \nonumber\\
    &= \begin{pmatrix} 0 & \Tilde{v}_{12} - i \hbar\mathbf{\Tilde{\Omega}}_{12} \cdot \boldsymbol{\hat \sigma} & (\Tilde{v}_{31} + i \hbar\mathbf{\Tilde{\Omega}}_{31} \cdot \boldsymbol{\hat \sigma})e^{-i\frac{\Tilde{\theta}_\text{so}^{(31)}}{2}\boldsymbol{\Tilde{\omega}}_{31} \cdot \boldsymbol{\hat \sigma}} \\
    h.c.  & 0 & (\Tilde{v}_{23} - i \hbar\mathbf{\Tilde{\Omega}}_{23} \cdot \boldsymbol{\hat \sigma})e^{-i\frac{\Tilde{\theta}_\text{so}^{(31)}}{2}\boldsymbol{\Tilde{\omega}}_{31} \cdot \boldsymbol{\hat \sigma}}\\ 
    h.c. & h.c. & 0\end{pmatrix} = \begin{pmatrix} 0 & \Tilde{v}_{12} - i \hbar\mathbf{\Tilde{\Omega}}_{12} \cdot \boldsymbol{\hat \sigma} & v_{31} \\
    h.c.  & 0 & \hat{v}_{23} - i \hbar\mathbf{\hat{\Omega}}_{23} \cdot \boldsymbol{\hat \sigma} \\ 
    h.c. & h.c. & 0\end{pmatrix},
    \label{eq:H3b3step2}
\end{align}
where we utilized the identity $e^{ia\boldsymbol{n \cdot \sigma}} = \cos(a) + i\sin(a)\boldsymbol{n\cdot\sigma}$, and introduced the parameters
\begin{subequations}
\begin{align}
    v_{31} &= \Tilde{v}_{31}\cos\left(\frac{\Tilde{\theta}_\text{so}^{(31)}}{2}\right) + \hbar|\boldsymbol{\Tilde{\Omega}}_{31}| \sin\left(\frac{\Tilde{\theta}_\text{so}^{(31)}}{2} \right) = \sqrt{\Tilde{v}_{31}^2 + \hbar^2|\boldsymbol{\Tilde{\Omega}}_{31}|^2},\\
    \hat{v}_{23} &= \Tilde{v}_{23}\cos\left(\frac{\Tilde{\theta}_\text{so}^{(31)}}{2}\right)  -  \hbar\boldsymbol{\Tilde{\Omega}}_{23} \cdot \boldsymbol{\Tilde{\omega}}_{31}\sin\left(\frac{\Tilde{\theta}_\text{so}^{(31)}}{2}\right) = \frac{\Tilde{v}_{23}\Tilde{v}_{31} - \hbar^2 \boldsymbol{\Tilde{\Omega}}_{23} \cdot \boldsymbol{\Tilde{\Omega}}_{31}}{\sqrt{\Tilde{v}_{31}^2 + \hbar^2|\boldsymbol{\Tilde{\Omega}}_{31}|^2}},\\
    \boldsymbol{\hat{\Omega}}_{23} &= \frac{1}{\hbar}\left((\Tilde{v}_{23}\boldsymbol{\Tilde{\omega}}_{31} + \hbar\mathbf{\Tilde{\Omega}}_{23}\cross \boldsymbol{\Tilde{\omega}}_{31})\sin\left(\frac{\Tilde{\theta}_\text{so}^{(31)}}{2}\right) + \hbar\mathbf{\Tilde{\Omega}}_{23} \cos\left(\frac{\Tilde{\theta}_\text{so}^{(31)}}{2}\right)\right) = \frac{\tilde{v}_{31}\boldsymbol{\Tilde{\Omega}}_{23} + \Tilde{v}_{23} \boldsymbol{\Tilde{\Omega}}_{31} + \hbar \boldsymbol{\Tilde{\Omega}}_{23}\cross \boldsymbol{\Tilde{\Omega}}_{31}}{\sqrt{\Tilde{v}_{31}^2 + \hbar^2|\boldsymbol{\Tilde{\Omega}}_{31}|^2}}.
\end{align}
\end{subequations} 
The pseudospin is indeed conserved upon tunneling between the third and the first dot in the new gauge, as $v_{31}$ appearing in the top right block of the matrix on the right hand side of Eq.~\eqref{eq:H3b3step2} is a scalar. 

Next, we eliminate the pseudospin-non-conserving tunneling between QD2 and QD3 with a rotation of the quantisation axes on the second dot around $\boldsymbol{\hat{\omega}}_{23} = \boldsymbol{\hat{\Omega}}_{23}/|\boldsymbol{\hat{\Omega}}_{23}|$ with an angle $\hat{\theta}_\text{so}^{(23)} = 2\arctan \frac{\hbar|\boldsymbol{\hat{\Omega}}_{23}|}{\hat{v}_{23}}$ in the clockwise (negative) direction. The corresponding unitary transformation is $U_2 = e^{-i\frac{\hat{\theta}_\text{so}^{(23)}}{2} \boldsymbol{\hat{\omega}}_{23} \cdot \boldsymbol{\hat \sigma}\otimes \tau_2}$. We denote the new basis as:
\begin{equation}
    \ket*{k, \Uparrow} = U_2\ket*{k, \hat \Uparrow}, \ket*{k, \Downarrow} = U_2\ket*{k, \hat \Downarrow},
\end{equation}
similarly to the previous case. The transformation of the matrix form of the tunneling Hamiltonian is analogous to the previous one. This means that the tunneling between QD1 and QD3 remains invariant, the tunneling between QD2 and QD3 become pseudospin-conserving and the last tunneling term between QD1 and QD2 transforms non-trivially. Written out with the Pauli operators $\boldsymbol{\sigma}$ defined in the new gauge:
\begin{align}
    H_\textrm{tun}& + H_\textrm{stun} = \nonumber\\
    &= \begin{pmatrix} 0 & (\Tilde{v}_{12} - i \hbar\mathbf{\Tilde{\Omega}}_{12} \cdot \boldsymbol{\sigma})e^{-i\frac{\hat{\theta}_\text{so}^{(23)}}{2}\boldsymbol{\hat{\omega}}_{23} \cdot \boldsymbol{\sigma}} & v_{31} \\
    h.c.  & 0 & e^{i\frac{\hat{\theta}_\text{so}^{(23)}}{2}\boldsymbol{\hat{\omega}}_{23} \cdot \boldsymbol{\sigma}}(\hat{v}_{23} - i \hbar\boldsymbol{\hat{\Omega}}_{23} \cdot \boldsymbol{\sigma})\\ 
    h.c. & h.c. & 0\end{pmatrix} = \begin{pmatrix} 0 & v_{12} - i \hbar\mathbf{\Omega}_{12} \cdot \boldsymbol{\sigma} & v_{31}\\
    h.c.  & 0 & v_{23} \\ 
    h.c. & h.c. & 0\end{pmatrix},
\end{align}
with the new parameters:
\begin{subequations}
\begin{align}
    v_{12} =& \Tilde{v}_{12}\cos\left(\frac{\hat{\theta}_\text{so}^{(23)}}{2}\right)- \sin\left(\frac{\hat{\theta}_\text{so}^{(23)}}{2}\right)\hbar\mathbf{\Tilde{\Omega}}_{12} \cdot \boldsymbol{\hat{\omega}}_{23} = \frac{\Tilde{v}_{12}\hat{v}_{23} - \hbar^2 \boldsymbol{\Tilde{\Omega}}_{12} \cdot \boldsymbol{\hat{\Omega}}_{23}}{\sqrt{\tilde{v}_{23}^2 + \hbar^2|\boldsymbol{\Tilde{\Omega}}_{23}|^2}} = \nonumber\\
    &=\frac{\Tilde{v}_{12}\Tilde{v}_{23}\Tilde{v}_{31} - \Tilde{v}_{12}\hbar^2 \boldsymbol{\Tilde{\Omega}}_{23} \cdot \boldsymbol{\Tilde{\Omega}}_{31} - \Tilde{v}_{31}\hbar^2 \boldsymbol{\Tilde{\Omega}}_{12} \cdot \boldsymbol{\tilde{\Omega}}_{23} - \Tilde{v}_{23} \hbar^2 \boldsymbol{\Tilde{\Omega}}_{12} \cdot \boldsymbol{\Tilde{\Omega}}_{31} - \hbar^3 \boldsymbol{\Tilde{\Omega}}_{12} \cdot (\boldsymbol{\Tilde{\Omega}}_{23}\cross \boldsymbol{\Tilde{\Omega}}_{31})}{\sqrt{(\tilde{v}_{23}^2 + \hbar^2|\boldsymbol{\Tilde{\Omega}}_{23}|^2)(\Tilde{v}_{31}^2 + \hbar^2|\boldsymbol{\Tilde{\Omega}}_{31}|^2})},\\
    v_{23} =& \Tilde{v}_{12}\cos\left(\frac{\hat{\theta}_\text{so}^{(23)}}{2}\right)- \sin\left(\frac{\hat{\theta}_\text{so}^{(23)}}{2}\right)\hbar\mathbf{\Tilde{\Omega}}_{12} \cdot \boldsymbol{\hat{\omega}}_{23}= \sqrt{\hat{v}_{23}^2 + \hbar^2|\boldsymbol{\hat{\Omega}}_{23}|^2} = \sqrt{\tilde{v}_{23}^2 + \hbar^2|\boldsymbol{\tilde{\Omega}}_{23}|^2},\\
    \boldsymbol{\Omega}_{12} =& \frac{1}{\hbar}\left((\Tilde{v}_{12}\boldsymbol{\hat{\omega}}_{23} +  \hbar\mathbf{\Tilde{\Omega}}_{12} \cross \boldsymbol{\hat{\omega}}_{23}) \sin\left(\frac{\hat{\theta}_\text{so}^{(23)}}{2}\right) + \cos\left(\frac{\hat{\theta}_\text{so}^{(23)}}{2}\right)\hbar\mathbf{\Tilde{\Omega}}_{12}\right) = \frac{\hat{v}_{23}\boldsymbol{\Tilde{\Omega}}_{12} + \Tilde{v}_{12} \boldsymbol{\hat{\Omega}}_{23} + \hbar \boldsymbol{\Tilde{\Omega}}_{12}\cross \boldsymbol{\hat{\Omega}}_{23}}{\sqrt{\Tilde{v}_{23}^2 + \hbar^2|\boldsymbol{\Tilde{\Omega}}_{23}|^2}} = \nonumber \\
    =& \frac{\Tilde{v}_{12}\Tilde{v}_{23}\boldsymbol{\Tilde{\Omega}}_{31} + \Tilde{v}_{31}\Tilde{v}_{12}\boldsymbol{\Tilde{\Omega}}_{23} + \Tilde{v}_{23}\Tilde{v}_{31}\boldsymbol{\Tilde{\Omega}}_{12}  + 
    \Tilde{v}_{12}(\hbar \boldsymbol{\Tilde{\Omega}}_{23}\cross \boldsymbol{\Tilde{\Omega}}_{31}) + 
    \Tilde{v}_{23}(\hbar \boldsymbol{\Tilde{\Omega}}_{12}\cross \boldsymbol{\Tilde{\Omega}}_{31}) + 
    \Tilde{v}_{31}(\hbar \boldsymbol{\Tilde{\Omega}}_{12}\cross \boldsymbol{\Tilde{\Omega}}_{23})}{\sqrt{(\tilde{v}_{23}^2 + \hbar^2|\boldsymbol{\Tilde{\Omega}}_{23}|^2)(\Tilde{v}_{31}^2 + \hbar^2|\boldsymbol{\Tilde{\Omega}}_{31}|^2})}
    - \nonumber\\ 
    &-\frac{\hbar^2 (\boldsymbol{\Tilde{\Omega}}_{23} \cdot \boldsymbol{\Tilde{\Omega}}_{31})\boldsymbol{\Tilde{\Omega}}_{12}  - (\boldsymbol{\Tilde{\Omega}}_{12} \cdot \boldsymbol{\Tilde{\Omega}}_{31}) \boldsymbol{\Tilde{\Omega}}_{23} + (\boldsymbol{\Tilde{\Omega}}_{12} \cdot \boldsymbol{\Tilde{\Omega}}_{23}) \boldsymbol{\Tilde{\Omega}}_{31}}{\sqrt{(\tilde{v}_{23}^2 + \hbar^2|\boldsymbol{\Tilde{\Omega}}_{23}|^2)(\Tilde{v}_{31}^2 + \hbar^2|\boldsymbol{\Tilde{\Omega}}_{31}|^2})}.
\end{align}
\end{subequations}
\twocolumngrid
To arrive to the form of Eq.~\eqref{eq:tunaftergauge}, one needs to rotate the basis with a global pseudospin transformation to rotate the $\boldsymbol{\Omega}_{12}$ vector to the $z$ direction. Therefore the $\Omega$ parameter in Eq.~\eqref{eq:stunaftergauge} reads as $\Omega = |\boldsymbol{\Omega}_{12}|$.

\section{Gauge transformation of the Zeeman term}
\label{app:gaugeZeeman}

Let us calculate the effect of a local basis transformation on the Zeeman term of the Hamiltonian. 
The local Zeeman term of a dot can be written as
\begin{equation}
    H_\text{Zeeman} = \frac{1}{2}\mu_\textrm{B} \boldsymbol{\tilde{\sigma}} \cdot \boldsymbol{\tilde{g} B},
\end{equation}
as given in Eq.~\eqref{eq:zeeman}. Consider the gauge transformation $U = e^{-i\frac{\theta_\text{so}}{2}\boldsymbol{n \cdot \tilde \sigma}}$, similar to the transformations in the previous appendix. 
Then, the Pauli operators defined with the new basis are related to the original ones by the formula
\begin{equation}
    \tilde{\boldsymbol{\sigma}} = U^\dagger \boldsymbol{\sigma} U = \boldsymbol{R_n}(\theta_\text{so})\boldsymbol{\sigma},
\end{equation}
where we introduced $\boldsymbol{R_n}(\theta_\text{so})$ as a 3$\cross$3 rotation matrix around $\boldsymbol{n}$ with angle $\theta_\text{so}$. Then the Zeeman term expressed using the Pauli operators of the new basis is
\begin{align}
    H_\text{Zeeman} &= \frac{1}{2}\mu_\textrm{B} (\boldsymbol{R_n}(\theta_\text{so})\boldsymbol{\sigma}) \cdot \boldsymbol{\tilde{g} B} \nonumber\\
    &= \frac{1}{2}\mu_\textrm{B} \boldsymbol{\sigma}\cdot (\boldsymbol{R_n}^{-1}(\theta_\text{so}) \boldsymbol{\tilde{g}}) \boldsymbol{B} \nonumber\\
    &= \frac{1}{2}\mu_\textrm{B} \boldsymbol{\sigma}\cdot \boldsymbol{ g B},
\end{align}
where we introduced the new $g$-tensor $\boldsymbol{g} = \boldsymbol{R_n}^{-1}(\theta_\text{so})\boldsymbol{\tilde{g}}$. 

Using the above considerations, we express the local $g$-tensors of the three-dot loop after the gauge transformation considered in the previous appendix:
\begin{subequations}
\begin{align}
   \boldsymbol{g}_1  &= \tilde{\boldsymbol{g}}_1,\\
   \boldsymbol{g}_2 &= \boldsymbol{R}_{\hat{\boldsymbol{\omega}}_{23}}^{-1}(\hat{\theta}_\text{so}^{(23)})\tilde{\boldsymbol{g}}_2,\\
   \boldsymbol{g}_3 &= \boldsymbol{R}_{\tilde{\boldsymbol{\omega}}_{31}}^{-1}(\tilde{\theta}_\text{so}^{(31)})\tilde{\boldsymbol{g}}_3.
\end{align}    
\end{subequations}

\section{Qubit readout error due to temperature broadening of the Fermi-Dirac distribution}
\label{app:readout}

The Elzerman readout scheme \cite{Elzerman2004Single-shotDot} for quantum-dot spin qubits relies on spin-to-charge conversion and subsequent charge readout. 
In this scheme, spin-to-charge conversion is enabled by a nearby reservoir of particles and energy-selective tunneling between the dot and the reservoir.
In particular, the on-site energy of the dot is tuned such that the chemical potential of the reservoir is in between the two Zeeman-split sublevels of the spin qubit, such that tunneling to the reservoir is suppressed (enhanced) for the ground (excited) state of the spin qubit.
In turn, the absence or presence of the particle in the dot is measured by a sensitive charge sensor, e.g., a quantum point contact or a single-electron transistor. 

The reservoir used for spin-to-charge conversion has a finite temperature, hence the Fermi-Dirac distribution of the particles in the reservoir is thermally broadened. 
This implies a nonzero readout error which we characterize in what follows. 
In particular, we ask the following question: if the spin qubit occupies the excited state with probability $P_\text{e}$ right before readout, then what is the probability $P_\text{e}^{(\text{m})}$ of the inference that it was in the excited state?
The fact that $P_\text{e} \neq P_\text{e}^{(\text{m})}$ (see below) is interpreted as a thermally induced readout error, and it influences the contrast of the oscillations described in Secs.~\ref{sec:NABP_meas} and \ref{sec:parametrisation}.

In our model, we define the excited-state inference probability $P_\text{e}^{(\text{m})}$ as the probability that the particle has left the dot at least once during the charge sensing measurement.
We consider the case when charge sensing is done such that the chemical potential of the reservoir aligns with the average energy of the two Zeeman sublevels of the spin qubit.
To describe the tunneling events, we introduce the tunneling-out rates as
\begin{subequations}
\begin{eqnarray}
    \Gamma_\textrm{g} = \Gamma (1 -  f(\mu - \hbar \omega /2)),\\
    \Gamma_\textrm{e} = \Gamma (1 -  f(\mu + \hbar \omega /2)),
\end{eqnarray}
\end{subequations}
where $\Gamma$ is the bare tunneling amplitude between the dot and the reservoir, $f$ is the Fermi-Dirac distribution and $\hbar\omega$ is the splitting between the excited state and the ground state. 

Let us assume, that the system is in the excited (ground) state. Then the probability that it has not tunneled out after time $\tau_\textrm{r}$ is $e^{-\Gamma_\textrm{e}\tau_\textrm{r}}$ ($e^{-\Gamma_\textrm{g}\tau_\textrm{r}}$). Thus, the probability of jumping out from the excited (ground) state is $1 - e^{-\Gamma_\textrm{e}\tau_\textrm{r}}$ ($1 - e^{-\Gamma_\textrm{g}\tau_\textrm{r}}$).
Therefore, the inferred excited-state probability reads as
\begin{align}
    P^{(\textrm{m})}_\textrm{e}(P_{\textrm{e}}, \tau_\textrm{r}) &= (1 - e^{-\Gamma_\textrm{e}\tau_\textrm{r}})P_{\textrm{e}} + (1 - P_{\textrm{e}})(1 - e^{-\Gamma_\textrm{g}\tau_\textrm{r}})\nonumber\\
    &= (e^{-\Gamma_\textrm{g}\tau_\textrm{r}} - e^{-\Gamma_\textrm{e}\tau_\textrm{r}})P_{\textrm{e}} + (1 - e^{-\Gamma_\textrm{g}\tau_\textrm{r}}).
\end{align}
This implies that the contrast of the signal obtained from many Elzerman-type measurement runs is reduced by the factor 
\begin{equation}
    \tilde A (\omega, T, \Gamma, \tau_\textrm{r}) = e^{-\Gamma_\textrm{g}\tau_\textrm{r}} - e^{-\Gamma_\textrm{e}\tau_\textrm{r}}.
\end{equation}
To maximize the contrast at a given magnetic field, temperature, and tunneling rate, this factor $\tilde{A}$ should be maximized over the readout time $\tau_\text{r}$.
This yields a maximal contrast at an optimal readout time of 
\begin{equation}
    \tau_\textrm{r}^* = \frac{\ln (\Gamma_\textrm{g}/\Gamma_\textrm{e})}{\Gamma_\textrm{g} - \Gamma_\textrm{e}}.
\end{equation}
In this optimized setting, the contrast reduction factor reads:
\begin{align}
    A\left( \frac{\hbar\omega}{2k_\textrm{B}T} \right) &= \tilde A (\omega, T, \Gamma, \tau_\textrm{r}^*) \nonumber \\
    &= \exp\left(\frac{\Gamma_\textrm{g}\ln (\Gamma_\textrm{e}/\Gamma_\textrm{g})}{\Gamma_\textrm{g} - \Gamma_\textrm{e}}\right) - \exp\left(\frac{\Gamma_\textrm{e}\ln (\Gamma_\textrm{e}/\Gamma_\textrm{g})}{\Gamma_\textrm{g} - \Gamma_\textrm{e}}\right) \nonumber \\
    &=\exp\left(-\frac{\frac{\hbar \omega}{2k_\text{B}T}}{e^{\frac{\hbar \omega}{2k_\text{B}T}} - 1}\right) - \exp\left(-\frac{\frac{\hbar \omega}{2k_\text{B}T}}{1 - e^{-\frac{\hbar \omega}{2k_\text{B}T}}}\right).
\end{align}
Note that the maximum value does not depend on the tunneling amplitude $\Gamma$. The formula of the measured excited-state probability after all reads as
\begin{equation}
    P^{(\textrm{m})}_\textrm{e} = A\left( \frac{\hbar\omega}{2k_\textrm{B}T} \right)P_{\textrm{e}} + 1 - \exp\left(-\frac{\frac{\hbar \omega}{2k_\text{B}T}}{e^{\frac{\hbar \omega}{2k_\text{B}T}} - 1}\right). \label{eq:optimalized_measured_excited_state}
\end{equation}
The parameters $a^{(m)}$ and $b^{(m)}$ in Eq.~\eqref{eq:measurementoutcomeNloop} can be calculated by substituting $P_{1,\textrm{e},N}$ defined in Eq.~\eqref{eq:occupationNloop} into $P_{e}$ in Eq.~\eqref{eq:optimalized_measured_excited_state}. The exact formulas read as
\begin{subequations}
\begin{align}
    a^{(\textrm{m})} &= A\left( \frac{\hbar\omega}{2k_\textrm{B}T} \right) a (\theta_1, T, \omega_1) + 1 - \exp\left(-\frac{\frac{\hbar \omega}{2k_\text{B}T}}{e^{\frac{\hbar \omega}{2k_\text{B}T}} - 1}\right),\\
    b^{(\textrm{m})} &=A\left( \frac{\hbar\omega}{2k_\textrm{B}T} \right) b(\theta_1, T, \omega_1),
\end{align}    
\end{subequations}
where $a(\theta_1, T, \omega_1)$ and $b(\theta_1, T, \omega_1)$ are defined in Eqs.~\eqref{eqs:rawparameters}.

\twocolumngrid
\bibliography{references, arxive}
\end{document}